\documentclass[sn-aps,pdflatex]{sn-jnl}% referee option is meant for double line spacing

\jyear{2022}%

\theoremstyle{thmstyleone}%
\theoremstyle{thmstyletwo}%

\theoremstyle{thmstylethree}%

\usepackage[utf8]{inputenc}
\usepackage{graphicx}

\usepackage{geometry}
\usepackage{caption}
\usepackage{subcaption}

\title[SPAD Ising/Potts Solver]{CMOS-compatible Ising and Potts Annealing Using Single Photon Avalanche Diodes}

\author[1]{\fnm{William} \sur{Whitehead}}\email{williamwhitehead@ucsb.edu}

\author[1]{\fnm{Zachary} \sur{Nelson}}\email{znelson@ucsb.edu}

\author[1]{\fnm{Kerem} \sur{Y. Camsari}}\email{camsari@ucsb.edu}

\author*[1]{\fnm{Luke} \sur{Theogarajan}}\email{lusthe@ucsb.edu}

\affil*[1]{\orgdiv{Electrical \& Computer Engineering Dept.}, \orgname{University of California}, \orgaddress{\city{Santa Barbara},  \state{CA}, \postcode{93106}, \country{U.S.A.}}}

\abstract{Massively parallel annealing processors may offer superior performance for a wide range of sampling and optimization problems.  A key component dictating the size of these processors is the neuron update circuit, ideally implemented using special stochastic nanodevices. We leverage photon statistics using single photon avalanche diodes (SPADs) and temporal filtering to generate stochastic states. This method is a powerful alternative offering unique features not currently seen in annealing processors: the ability to continuously control the computational temperature and the seamless extension to the Potts model, a $n$-state generalization of the two-state Ising model. SPADs also offer a considerable practical advantage since they are readily manufacturable in current standard CMOS processes. As a first step towards realizing a CMOS SPAD-based annealer, we have designed Ising and Potts models driven by an array of discrete SPADs and show they accurately sample from their theoretical distributions.}

\begin{document}

\maketitle

\section{Introduction}

Resource-intensive computing tasks are often off-loaded to  specialized processors to accelerate execution.  One such task is simulated annealing \cite{original_simulated_annealing}, a probabilistic method of finding solutions to hard optimization problems occurring in a variety of fields such as chip floor planning \cite{simulated_annealing_chip_floorplanning} and logistics \cite{simulated_annealing_in_operations_research}.  These optimization problems typically require exploration over an exponentially large solution space in order to find acceptable answers, and simulated annealing prescribes an efficient probabilistic method of searching over this solution space but is computationally expensive.  While simulated annealing may take many forms, current processors dedicated to the acceleration of simulated annealing have exclusively focused on the Ising model \cite{Ising_formulations, review_of_hw_Ising_solvers}.  The Ising model is preferred due to its simplicity, enabling easy emulation in hardware, and universality, since it can represent any discrete optimization problem \cite{Ising_formulations}.  Each variable is usually mapped directly to its own emulating circuit in an annealing processor to achieve the highest possible computational speed and efficiency \cite{MTJ_for_pbits,2x30k_CMOS_annealer,compute_in_memory_CMOS_annealer,massively_parallel_sparse_Ising}.  This direct emulation is preferred over more sequential Ising model processor architectures \cite{series_parallel_RBM_in_FPGA, DEC_coprocessor_BM}  usually designed for machine learning purposes, since optimization problems require substantial computation on relatively small graphs. Scaling the Ising model to large networks requires efficient design of the emulation circuit, commonly referred to as the neuron or spin. 

There are multiple approaches to the problem of constructing simulated annealers  \cite{review_of_hw_Ising_solvers}, but no clear best method has emerged so far.  Computationally, Ising machines require the multiply and accumulate operation for calculating the Hamiltonian (see equation \ref{eqn:E_Ising}), followed by calculating the probability of flipping the state, usually via the tangent hyperbolic function and a uniform random number generator. Proposed designs can be placed in several categories, including FPGA implementations, various ASICs, and novel device designs. Novel device designs leverage specific stochastic microelectronic devices to calculate the probability of a state flip by one small device.  A prime example in this category is the use of magnetic tunnel junctions, which stochastically exist in one of two magnetization states with a probability tunable by a bias input \cite{MTJ_for_pbits}.  Three-terminal memristors have been designed for similar operation \cite{memristor_Ising_machine}, and a variety of devices have been used in other ways \cite{memristor_weight_noise_hopfield_Ising, coherent_Ising_perspective, atomic_boltzmann}.  ASIC designs have been proposed based on both digital and analog methods.  Digital CMOS annealers have reduced the size of individual neuron circuits through optimized digital design \cite{efficient_tanh_example} and trading compactness for precision, such as reducing arithmetic to only a few bits and replacing hyperbolic tangent and random number generation with simpler circuits demonstrating random behavior \cite{2x30k_CMOS_annealer, compute_in_memory_CMOS_annealer}.  Analog computation is naturally able to pack complex functions such as hyperbolic tangent activation into simple circuits, but many analog circuit designs search for solutions using mean-field behavior rather than Gibbs sampling \cite{analog_bistable_Ising_machine, analog_LC_oscillator_Ising}.  FPGA Ising machine implementations often take advantage of the scale of modern FPGAs and utilize efficient graph implementation techniques, such as sparsification \cite{massively_parallel_sparse_Ising}.  

The proposed SPAD-based design showcases a novel stochastic computational device offering the ability to be readily integrated with current CMOS technology. We present a method of using single-photon avalanche diodes (SPADs) as neuron circuits in simulated annealing processors, with a number of features unavailable in existing designs, such as the ability to simulate the Potts model in addition to the Ising model. We show results from a hardware realization of this approach and demonstrate its ability to reproduce the theoretical distribution via Gibbs sampling. The design consists of 8 Ising neurons or 4 Potts neurons implemented using a 4x4 silicon photon multiplier array (each element is an array of SPADs), see Figure 1. The variable rate SPAD circuit (VRSC) is the central element enabling the use of the SPAD as a stochastic element in an Ising or Potts model. Briefly, the random current  pulses of the SPAD are converted into an amplitude via a filtering circuit and compared to a reference proportional to the computational energy of the network. The resulting pulse rate is proportional to the exponential of the negative computational energy (see section \ref{sec:TOP}). The rate is subsequently converted into a state using a latching circuit yielding a stochastic node following Boltzmann statistics.  In the next few sections following a brief introduction to the Ising and Potts models, we describe the architecture, theory of operation and discuss experimental results of our SPAD enabled Ising/Potts solvers.

\begin{figure}[!t]
\begin{center}
    \begin{subfigure}[b]{4.2in}
        \centering
        \includegraphics[width=\textwidth]{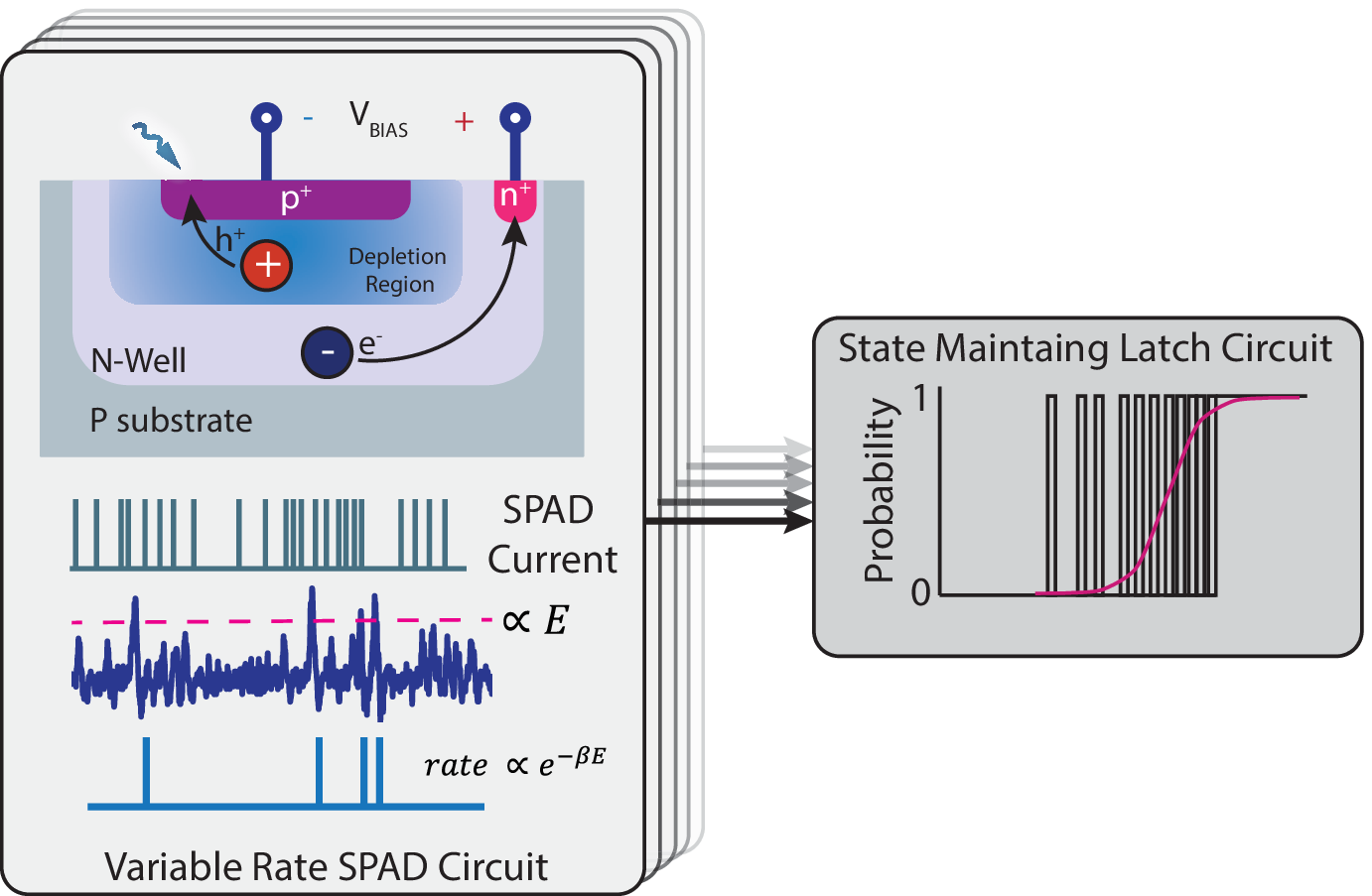}
        \caption{Illustrative operation of the Variable Rate SPAD-based Neuron}
        \label{fig:concept}
    \end{subfigure}
    \hfill
     \begin{subfigure}[b]{2in}
        \centering
        \includegraphics[width=2in]{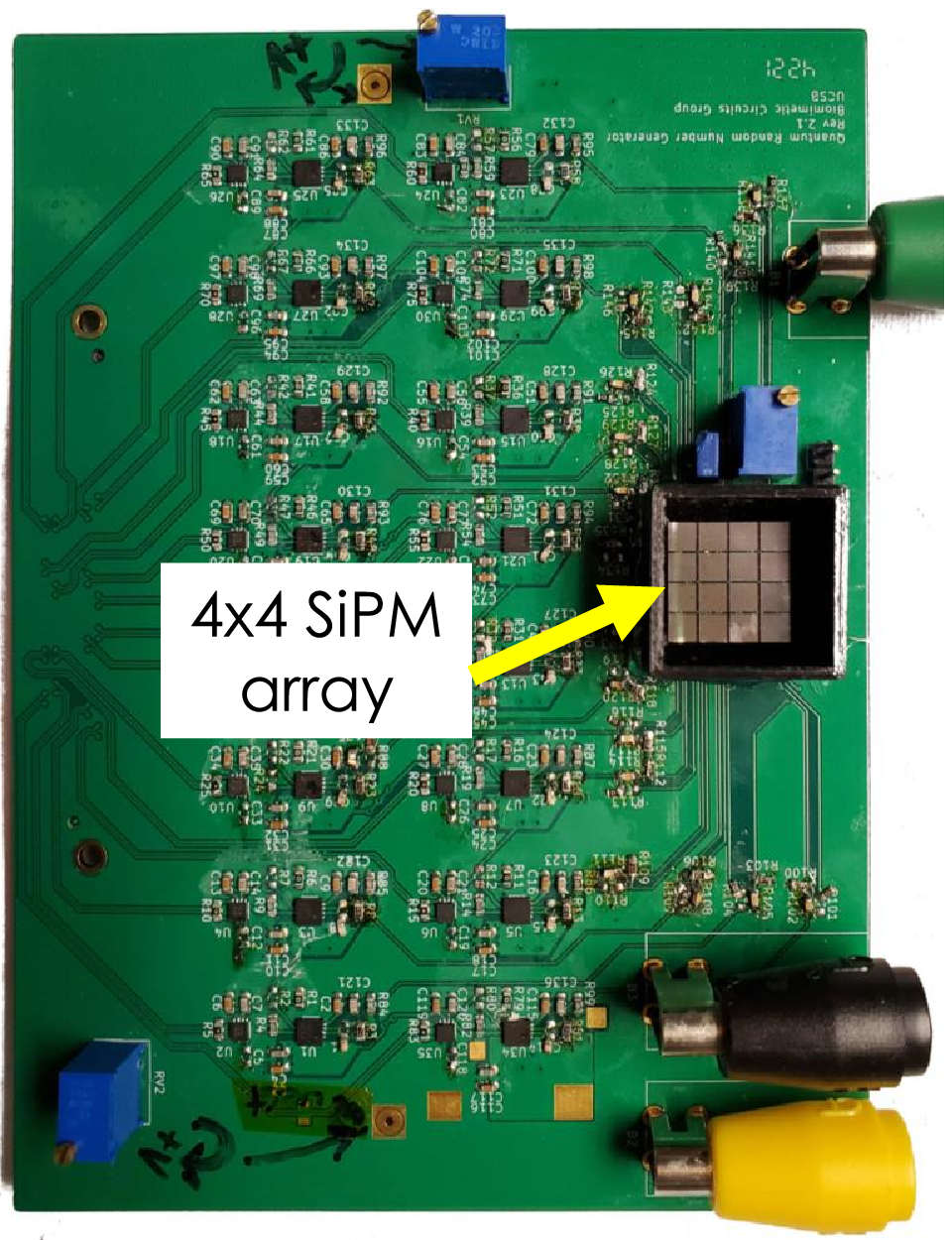}
        \caption{PCB based Variable Rate SPAD circuits}
        \label{fig:test_pcb}
    \end{subfigure}
    
    \vspace{0.2in}

    \begin{subfigure}[b]{6.5in}
        \centering
        \includegraphics[width=6.5in]{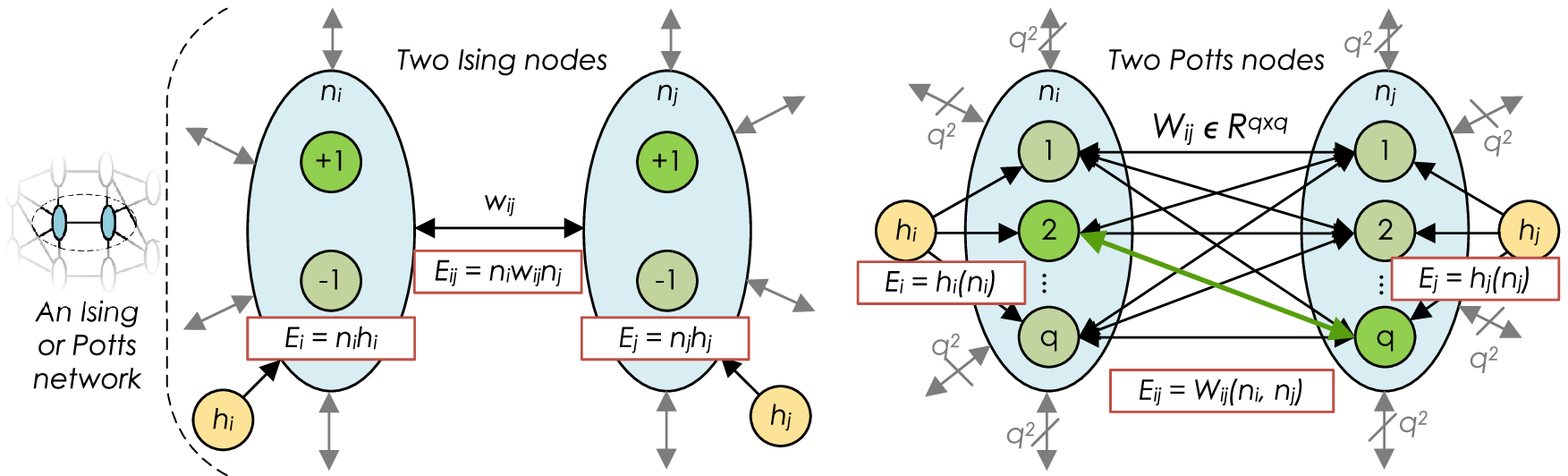}
        \caption{Ising and Potts models}
        \label{fig:Ising_potts_explenation}
    \end{subfigure}
\end{center}
\caption 
{ \label{fig:overview}
Overview of CMOS-compatible Ising and Potts annealing using SPADs.  a) Energy signals control the rate of Poisson pulses from circuits based on CMOS SPADs, which are coupled to a latch to form the output state of an Ising or Potts neuron.  b) A proof-of-concept is demonstrated using discrete components, and is found to accurately replicate theoretical distributions. c) Architecture of Ising and Potts models.  In an Ising model, a binary variable $n_i$ contributes to the energy of the model through a linear term set by $h_i$ and through quadratic terms set by weights $w_{ij}$.  In the Potts model each variable may take one of $q$ states, each with an independent meaning, requiring the energy to be expressed as the selection of a weight rather than as a product of a weight.} 
\end{figure}

\subsection{Implementing Ising and Potts Models} 

Both Ising and Potts models allow the mapping of the NP complete problem class to a hardware structure capable of finding approximate solutions in polynomial time \cite{Ising_formulations, review_of_hw_Ising_solvers, ising_history, potts_history}. A major difference between the Ising and Potts models is the representational ability of each neuron or spin. The Ising model utilizes binary valued neuron as opposed to the Potts model, which can be in one of $q$ possible states. Typically, the Potts neurons are represented by a one-hot encoded $q$-state vector. A central contribution of this work is to enable hardware realizations of the Potts model. The Potts model provides a more natural representation of many optimization problems making it more efficient in finding solutions \cite{potts_superiority, potts_vs_Ising_natural_representation, binary_vs_onehot_encoding}.  Graph coloring with $q$ colors can be mapped directly to Potts model neurons with $q$ states, but mapping to the Ising model requires splitting each problem node across $q$ different Ising neurons \cite{potts_graph_coloring}. However, this increase in representational ability comes at the expense of requiring more weights. Ising model requires single weights between each pair of neurons. Potts neurons on the other hand require $(q-1)^2$ weights between each pair of neurons. Unlike the Ising model, where the weights encode the degree of (anti)correlation, each weight in the Potts model encodes the cost of each of $q^2$ possible joint neuron configurations.

The Ising and Potts models are best understood in the language of statistical mechanics \cite{ising_history,potts_history}.  The collection of variables $\{n_1,...n_N\}$ in a system form a joint state $s$ with energy $E$. This energy depends on the interaction of the individual neuron states $n_i,n_j$ through the model weights $w_{ij}$.  For the Ising and Potts models, the energy of a state is defined as 
\begin{equation}
    E_{Ising}(s = \{n_1, ...n_N\}) =-\left( \frac{1}{2}\sum_{i=1}^{N}\sum_{j=1}^{N} w_{ij}n_in_j + \sum_{i=1}^N h_in_i\right)
    \label{eqn:E_Ising}
\end{equation}
\begin{equation}
    E_{Potts}(s = \{n_1, ...n_N\}) =-\left(\frac{1}{2}\sum_{i=1}^{N}\sum_{j=1}^{N} W_{i,j}(n_i,n_j) + \sum_{i=1}^N h_i(n_i)\right)
    \label{eqn:E_potts}
\end{equation}
The energy of the Potts model is also often expressed using the delta function formulation; the two forms are equivalent.  As a statistical mechanical model, the energy determines the probability of each joint state according to the Boltzmann distribution.  While optimization algorithms based on Ising and Potts models need not produce a Boltzmann distribution, most are designed to do so in order to take advantage of the understanding conferred by statistical mechanical theory.  In particular, when the energy of either an Ising or Potts model is the cost function of an optimization problem, the optimal solution is known to be the most likely solution (even if that probability is vanishing small).  In addition, the Boltzmann distribution provides the concept of temperature, which is essential for the simulated annealing process.  As with annealing of physical materials, slowly lowering the temperature allows the Ising or Potts model to settle into low energy configurations; as annealing to zero temperature is drawn out, the probability of the optimal state tends towards one \cite{simulated_annealing_book}.  In order for an Ising or Potts model to correctly sample from the Boltzmann distribution, state updates must conform to Gibbs sampling \cite{original_gibbs_sampler}.  Gibbs sampling dictates that if neurons are updated sequentially (no linked neurons are updated at the same time) and each update is according to the conditional Boltzmann distribution, then the joint distribution will be the Boltzmann distribution.  Alternative methods of finding solutions on Ising and Potts models that do not yield samples from the Boltzmann distribution but are nonetheless effective in some cases include mean-field approximations \cite{potts_superiority}, deterministic evolution as in Hopfield neural networks \cite{analog_bistable_Ising_machine}, and inexact probabilistic updating \cite{2x30k_CMOS_annealer,compute_in_memory_CMOS_annealer}.  Although these annealing processors do not perform exact Gibbs sampling, their neuron update circuits consist of nearly the same set of computational stages as those required for Gibbs sampling.

\begin{figure}[!t]
\begin{center}
    \begin{subfigure}[b]{6in}
        \centering
        \includegraphics[width=5in]{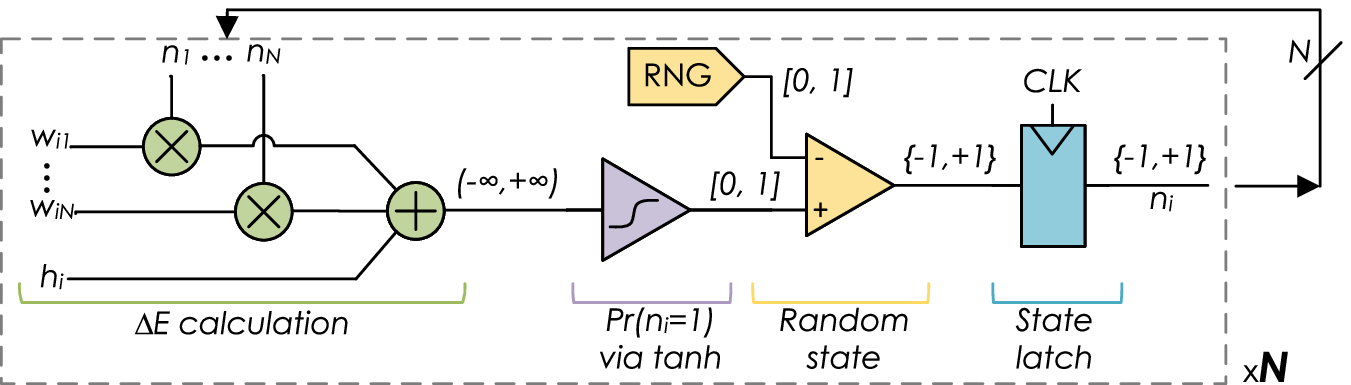}
        \caption{Signal flow for sampling from an Ising model}
        \label{fig:ising_comp_flow}
    \end{subfigure}
    
    \vspace{0.2in}
    
    \begin{subfigure}[b]{6in}
        \centering
        \includegraphics[width=5in]{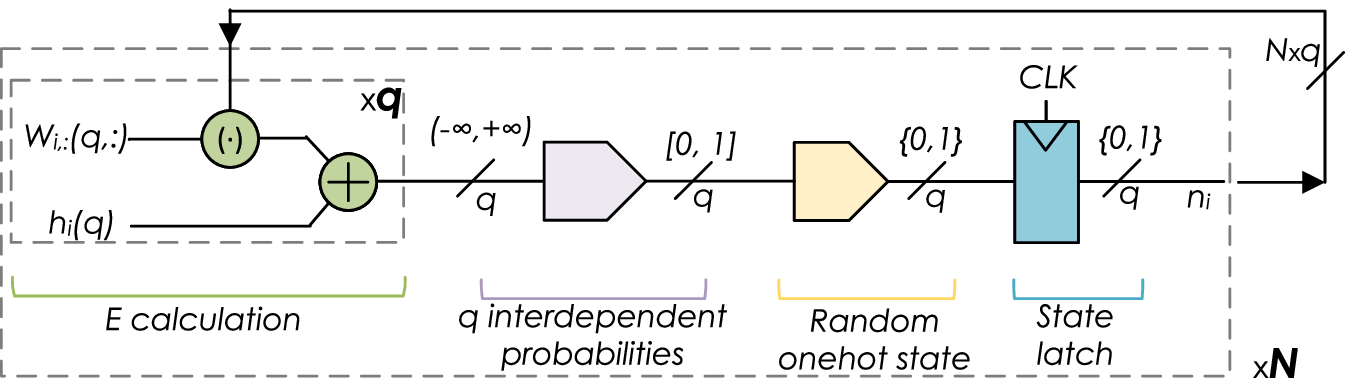}
        \caption{Signal flow for sampling from a Potts model}
        \label{fig:potts_comp_flow}
    \end{subfigure}
\end{center}
\caption 
{ \label{fig:standard_Ising_update}
Typical computational steps for simulating Ising or Potts models. Ising model computation can be broken into simple steps that may be readily performed by a number of hardware designs.  Many annealing processor designs (including the proposed design) perform these calculations without using identifiable units for each of these computational steps, but are ultimately performing the same calculations. While computation in the Potts model is structurally very similar, probability calculation and state selection become much more complicated since they occur in a one-hot variable space.} 
\end{figure}

The popularity of the Ising model stems from its relatively simple, hardware friendly, feed-forward structure.  During an update there are only two possibilities ($\pm 1$) to choose from, and therefore it is sufficient to only consider the difference in energy $\Delta E$ between the two states, and the probability of being in one state or the other can be expressed compactly using the hyperbolic tangent function:
\begin{equation}
    P(n_i = +1) = \frac{1+tanh(-\Delta E/2T)}{2}
    \label{eqn:tanh}
\end{equation}
Randomly generating a $\pm 1$ state according to this distribution can be accomplished by simply comparing the calculated probability to a uniformly generated random number and latching the result. For exact sampling, conditionally dependent neurons must be updated on different clock cycles. However, this constraint is often relaxed in order to achieve a higher level of parallelism.  This computational process, consisting of an energy sum, hyperbolic tangent activation, random number generation, comparison, and state latch is shown in figure (\ref{fig:standard_Ising_update}).  This design is a reference implementation of the Ising model which correctly samples from the Boltzmann distribution. In contrast, there is no equivalent design for Potts model neurons, since calculating a distribution and taking a random sample over many possibilities cannot be done so easily using feed-forward computational blocks.  The hardware resource consumption of even the Ising neuron reference design becomes problematic when scaling to designs with thousands of neurons, a limitation prompting a great deal of interest in building efficient hardware implementations of Ising machines.

\section{Results}

The single-photon avalanche diode (SPAD) is at first glance a natural candidate for addressing some of the shortcomings of existing designs.  SPADs are diodes biased beyond their reverse breakdown voltage, allowing amplification of single photons into detectable signals.  Under constant dim illumination the random arrival of photons makes a SPAD a source of random events, a feature that can be used for making true random number generators \cite{spad_qrng_1, spad_qrng_2}.  SPADs can also be currently integrated into CMOS with no process modifications, and are frequently designed into chips for imaging  \cite{1um_130nm_spad, spad_imager_review}.  The random behavior of a SPAD makes it appear to be another device around which compact neuron-update circuits can be built, similar to magnetic tunnel junctions and memristors, but with the added benefit of being implementable in a standard CMOS process.  However unlike MTJs and memristors, SPADs contribute random events, not a random state, and therefore must be used slightly differently.  SPADs could be incorporated simply as random number generators, by replacing the RNG block in the standard Ising neuron design with a SPAD random number generator \cite{spad_qrng_1, spad_qrng_2}.  However these random number generators require a lot of additional circuity, and from the standpoint of scaling down the area of individual neurons, this use of SPADs provides no benefits.  Instead we demonstrate a better method of using SPADs that turns the random events into an advantage and places the SPADs in a more central role.

% \begin{table}
% \begin{center}
% \begin{tabular}{l c c c c}
% Technology & CMOS & Model & Statistics & Temperature \\
% \hline
% This Work  & compatible & Ising/Potts & Boltzmann & Yes \\
% Magnetic Tunnel Junction \cite{MTJ_for_pbits} & not yet & Ising & Boltzmann & Yes \\
% Stochastic memristor \cite{memristor_Ising_machine} & not yet & Ising & Boltzmann & Yes \\
% Digital ASIC \cite{2x30k_CMOS_annealer} & Yes & Ising & non-Boltzmann & Yes \\
% Analog Discrete \cite{analog_bistable_Ising_machine} & compatible & Ising & None & No \\
% FPGA \cite{massively_parallel_sparse_Ising} & Yes & Ising & Boltzmann & No \\
% Bose-Einstein Condensate \cite{bose-einstein_condensate_potts} & No & Discrete XY & Boltzmann & Yes

% \end{tabular}
% \end{center}
% \caption 
% { \label{tab:comparison}
% Features offered by various annealing technologies.  Designs are identified by their compatibility with CMOS processes, what type of graph models they support, what kind of statistics they are designed to follow, and whether or not there is an option for continuously controlling the computational temperature.  Important features of an annealing processor such as size and graph connectivity are left out since they are not established by the state updating mechanism, which is the focus of this work.} 
% \end{table} 

Our proposed SPAD-based annealing design relies on two key design ideas, and provides a set of features unseen in any other annealing processor design.  The first design feature is the coupling of SPAD pulses to a latch, bridging the worlds of random events and random states.  The second is the construction of variable-rate SPAD circuits, which are SPAD-based circuits producing pulses (events) with a Poisson distribution and an input dependent mean frequency.  When combined, these two functions emulate a continuous-time Markov chain (CTMC), lending the name SPAD CTMC neurons.  A SPAD CTMC neuron may function just as a magnetic tunnel junction or three-terminal memristor: it has a stochastic output state whose distribution is controlled by an analog input.  In addition to presenting this functionality for Ising neurons, a SPAD CTMC neuron can also be constructed with more than two states, enabling the emulation of the Potts model with large $q$.  While annealing designs for the Potts model have been proposed before, none are as practical or as capable as the proposed SPAD-based design.  Hybrid approaches for annealing Potts models on Ising hardware have been described \cite{binary_vs_onehot_encoding,hybrid_potts_on_CIM, Potts_on_Ising_via_halfhot}, but introduce significant overhead and do not fundamentally operate as Potts models; a theoretical proposal to use Bose-Einstein condensate \cite{bose-einstein_condensate_potts} is a discretization of the XY model that is not capable of encoding typical optimization problems.  The SPAD CTMC neuron also allows for tuning the effective temperature of the Boltzmann distribution: temperature may be controlled by setting the illumination and bias conditions of the SPADs.  In the remainder of this paper we explain and demonstrate all of these features. 

\subsection{SPAD CTMC neuron architecture}

\begin{figure}[!t]
\begin{center}

    \hfill
    \begin{subfigure}[b]{3in}
        \centering
        \includegraphics[width=2.5in]{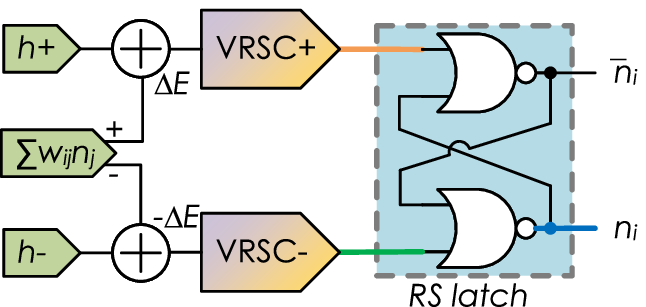}
        \caption{Proposed Ising node architecture}
        \label{fig:CTMC_Ising_node}
    \end{subfigure}
    \hfill
    \begin{subfigure}[b]{3in}
        \centering
        \includegraphics[width=3in]{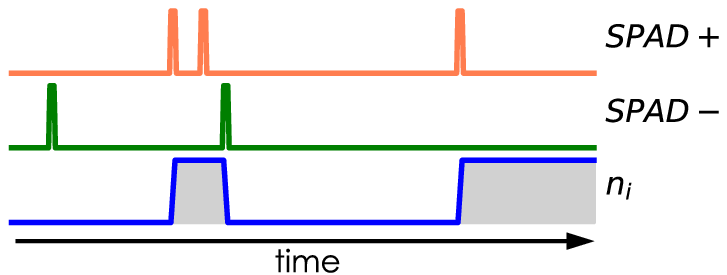}
        \caption{Latching in the Ising node}
    \end{subfigure}
    \hfill
    \hfill
    
    \vspace{0.1in}
    
    \hfill
    \begin{subfigure}[b]{3in}
        \centering
        \includegraphics[width=2.5in]{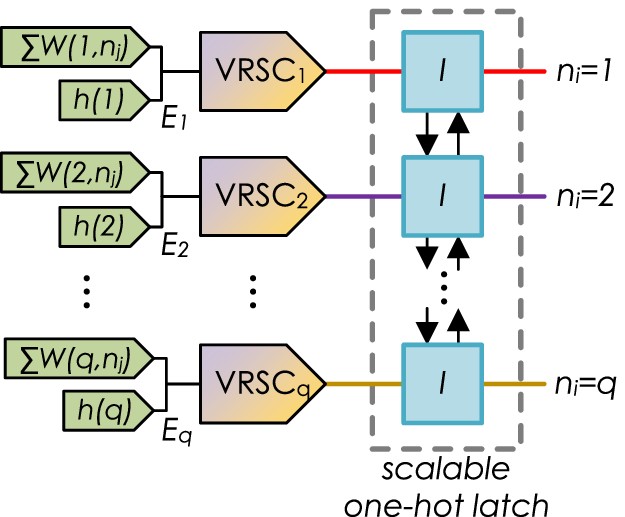}
        \caption{Proposed Potts node architecture}
        \label{fig:CTMC_potts_node}
    \end{subfigure}
    \hfill
    \begin{subfigure}[b]{3in}
        \centering
        \includegraphics[width=3in]{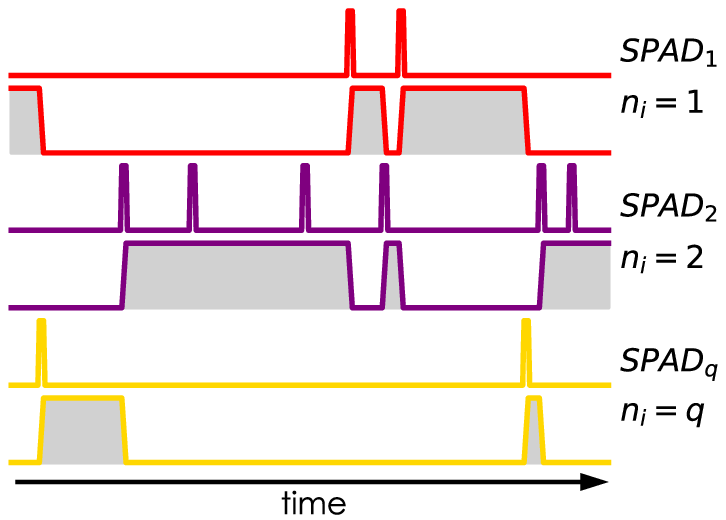}
        \caption{Latching in the Potts node}
    \end{subfigure}
    \hfill
    \hfill
    
    \vspace{0.1in}
    
    \hfill
    \begin{subfigure}[b]{0.45\textwidth}
        \centering
        \includegraphics[width=2.4in]{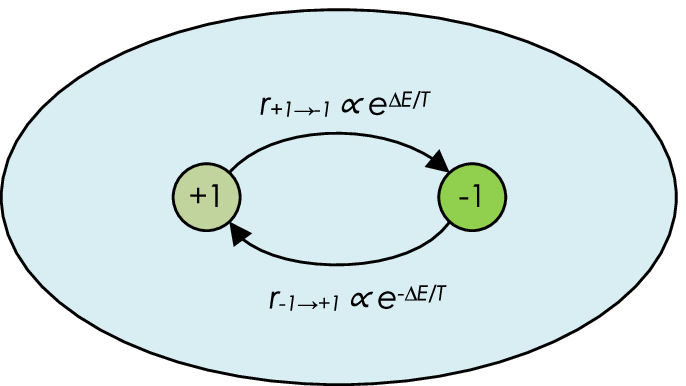}
        \caption{Ising node FSM}
    \end{subfigure}
    \hfill
    \begin{subfigure}[b]{0.45\textwidth}
        \centering
        \includegraphics[width=2.4in]{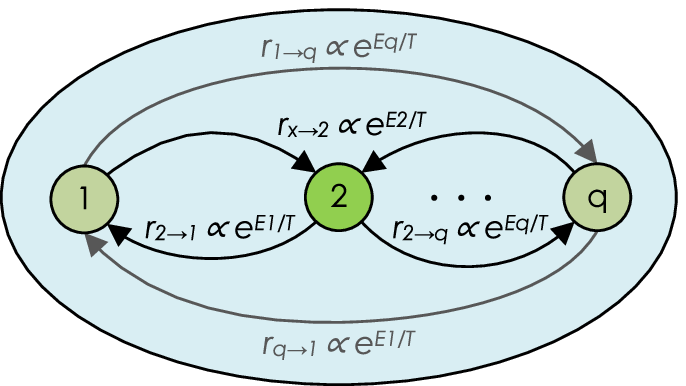}
        \caption{Potts node FSM}
    \end{subfigure}
    \hfill
    \hfill
    
\end{center}
\caption 
{ \label{fig:SPAD_CTMC_architecture}
Proposed circuits for Ising (a, b, e) and Potts (c, d, f) model computation.  Analog energy sums $E$ (which may be calculated in any way) control the rate at which variable-rate SPAD circuits (VRSCs) produce random events which in turn controls how much time the output latches spend in any one state.  The system is best modeled as continuous-time finite state Markov chains (e, f).  The rate at which the FSM transitions to a state (regardless of starting state) is the rate at which the associated SPAD circuit produces events.} 
\end{figure} 

The architecture of SPAD CTMC neurons is best understood if the design of variable-rate SPAD circuits is considered separately.  The operation of both Ising and Potts neurons starts with the calculation of conditional energies for each possible state, as shown in the figure \ref{fig:standard_Ising_update}.  The multiply-accumulate (MAC) calculation required for this step may be done in many ways but is most compactly performed in the current domain, either using digital weights \cite{Hopfield_digital_weights_current_sum} or memristive analog weights; since SPAD CTMC neurons are designed for CMOS integration, charge-trapped transistors (CTTs) \cite{CCT_CMOS_analog_synapse} would be a good choice since they are also CMOS devices. Once analog signals representing energy are calculated, they are used to control the variable-rate SPAD circuits.  Lower calculated energy (indicating a more favorable state) causes a variable-rate SPAD circuit to produce pulses more often.  At this stage the design of Ising and Potts neurons differentiates.

A neuron in the Ising model consists of two variable-rate SPAD circuits and an RS latch, configured as in figure \ref{fig:CTMC_Ising_node}.  When a SPAD circuit produces a pulse, the neuron transitions to a state corresponding to the respective SPAD.  The distribution of the neuron's state is controlled by the relative rate with which the two SPADs trigger.  If the SPADs trigger at the same rate, the neuron will spend equal time in the $+1$ and $-1$ states, regardless of the common mode event rate.  Changing the distribution requires adjusting the event rate of one or both SPADs.  Figure \ref{fig:CTMC_Ising_node} shows a symmetric solution in this regard, with the two SPADs adjusted in opposite directions, based on a single differential input sum representing the energy difference between the two states.  Natural variability in the SPAD event rate can be compensated for using a bias term, but does not present an overhead since it is a necessary term to encode optimization problems in the Ising model. If two biases are included, one for each variable-rate SPAD as in figure \ref{fig:CTMC_Ising_node}, the common-mode event rate may be controlled independently as well. This may or may not be needed in a design since small variations in the common mode rate between neurons in a network do not affect the statistical operation of the annealer. However, the overall rate must be set low enough such that neuron state updates have time to propagate to other neurons before the next event.  Alternative energy sum arrangements to achieve the Ising model are also possible. For example, arranging the weights as in the $q=2$ Potts model would obviously work, however the single differential energy sum method described here takes advantage of the simplifications possible in the Ising case, allowing more compact circuits than $q=2$ Potts model neurons.  

The circuit for the Potts model, shown in figure \ref{fig:CTMC_potts_node}, requires individual sums and SPADs for each possible state, and requires more complex latching circuitry (see supplementary material).  The behavior is similar to the Ising model where each SPAD attracts the neuron state to its index when it produces an event. The probability of each state is controlled by the relative event rates of the SPADs, which in turn are controlled by energy sums. The event-based nature of the SPAD seamlessly lends itself to constructing this design.  Since a simple RS-latch will no longer maintain a one-hot neuron state between SPAD events, a new latching design is required.    Similar in spirit to winner-take-all circuits \cite{analog_winner_take_all}, one of the $q$ latch outputs needs to be active at any time, and needs to indicate which SPAD circuit most recently pulsed.  A potential design outlined in figure \ref{fig:CTMC_potts_node} uses a lateral message-passing design where any element that is 'set' will simultaneously pass a 'reset' signal to all other elements through a daisy-chain.  While this design creates a delay proportional to the number of stages through which the message must be passed, it has the benefit of having $O(N)$ complexity and may be partitioned.  A string of 20 connected elements could for instance be arranged as a single $q=20$ neuron, or as four $q=5$ neurons, allowing for versatile hardware.  A logical implementation of this latch is available in the supplementary material. Using this SPAD CTMC neuron architecture, it is possible to build an annealing processor with configurable neuron dimensionality $q$ with similar complexity as Ising annealing processors. The proposed approach is widely applicable beyond SPAD based neurons.

\subsection{Theory of Operation \label{sec:TOP}}

The operation of each individual neuron is best modeled as a continuous-time Markov chain, since the latching circuit in SPAD CTMC neurons is a finite state machine transitioning randomly each time a variable-rate SPAD pulses.  Each neuron is naturally modeled as a CTMC since Boltzmann machines are known to be Markov random fields. While the following analysis is tailored to the Potts notation it is equally applicable to the operation of Ising nodes.  As long as the pulse timing is memoryless, then the pulse rates are a sufficient description of how the CTMC neurons transition between states.  The transition rate from state $i$ to state $j$ is the event rate of SPAD $j$, since each SPAD attracts the neuron state to itself regardless of the prior neuron state. The balance equation, which equates the rate $r_i$ at which a state is entered to the rate $\sum_{j \ne i}r_j$ at which a state is left, is used to find the probability $P_i$ of state $i$ in this CTMC and can be written as follows: 
\begin{equation}
    P_i\sum_{j \ne i}r_j = (1-P_i)r_i
\end{equation}
Which is easily manipulated into the form,
\begin{equation}
    P_i = {r_i \over{\sum_{j=1}^q r_j}}
\end{equation}

The probability of each state in the CTMC now has a similar form to that required by Gibbs sampling.  In Gibbs sampling  each time the neuron $n_i$ updates it must select from its allowed states according to the energy $E_Q$ of each state $Q$, conditioned on the rest of the network $\{n_j: j \ne i\} $: 
\begin{equation}
    P(n_i=Q) = {{e^{-E_Q/T}\over{\sum_{Q'}e^{-E_{Q'}/T}}}}
\end{equation}

Therefore, in order for the CTMC to sample from the Boltzmann distribution, the transition rate $r_Q$ must be equal to the exponential of the state's energy,

\begin{equation}
    r_Q = e^{-E_Q/T}
    \label{eqn:spad_requirement}
\end{equation}

If the event rate of a SPAD is an exponential function of some energy-representing control input, the Ising or Potts machine will correctly sample from the Boltzmann distribution.  The remaining challenge is constructing a variable rate SPAD circuit achieving this exponential transfer function.

\begin{figure}[t!]
\begin{center}

    \begin{subfigure}[b]{0.45\textwidth}
        \centering
        \includegraphics[width=3.0in]{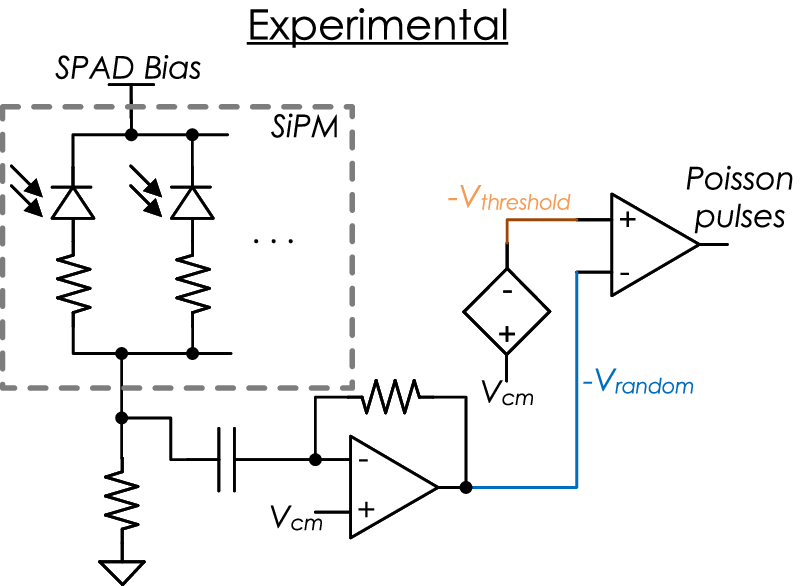}
        \caption{Experimental variable-rate SPAD circuit}
        \label{subfig:exp_circuit}
    \end{subfigure}
    \hfill
    \begin{subfigure}[b]{0.45\textwidth}
        \centering
        \includegraphics[width=3.0in]{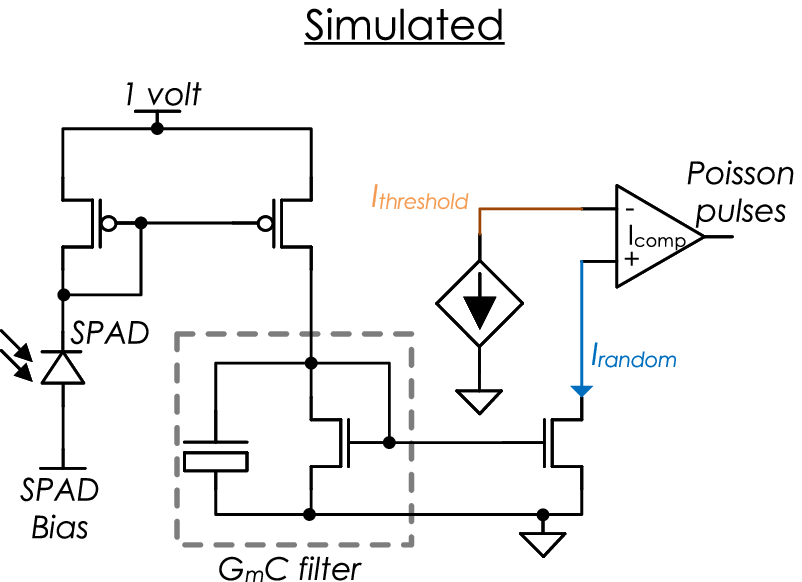}
        \caption{Simulated variable-rate SPAD circuit}
        \label{subfig:sim_circuit}
    \end{subfigure}
    
    \vspace{0.1in}

    \begin{subfigure}{0.45\textwidth}
      \centering
        \includegraphics[width=\linewidth]{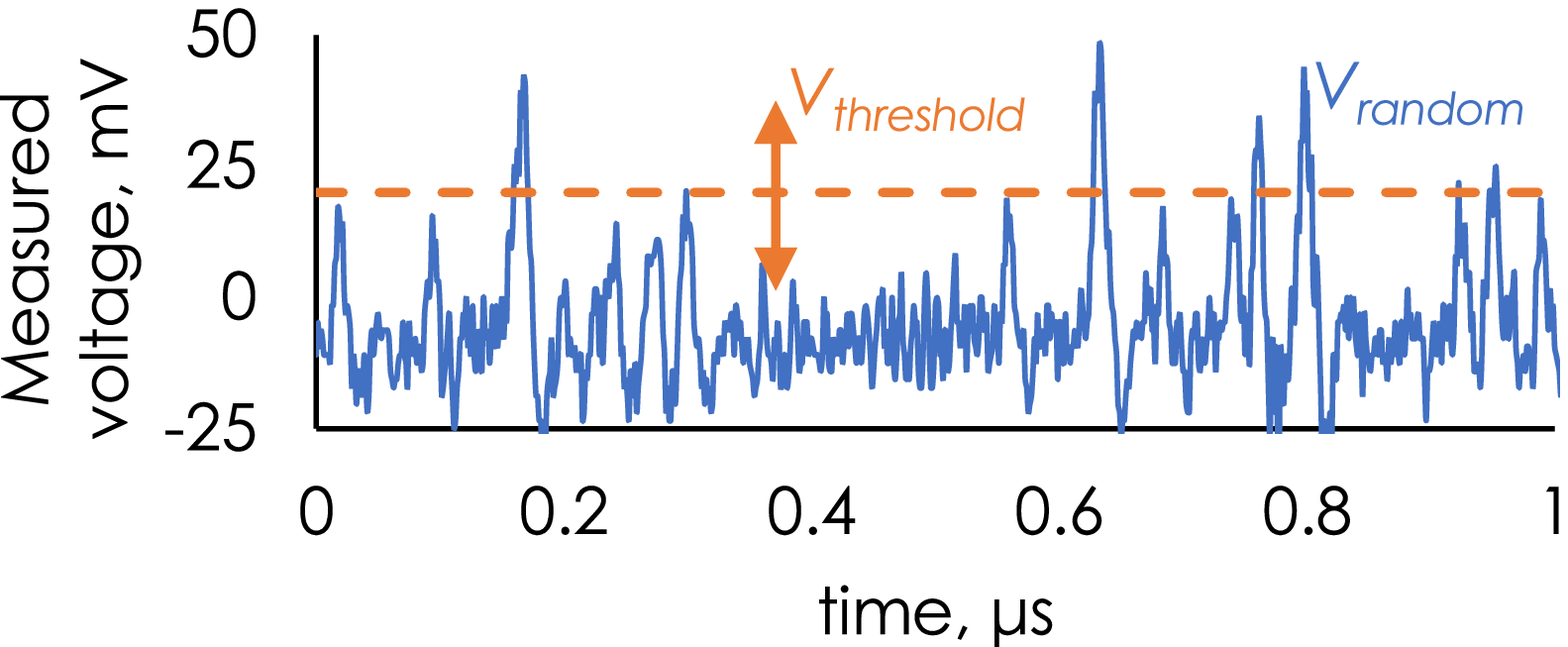}
        \caption{Measured random voltage}
        \label{subfig:rand_voltage}
    \end{subfigure}
   \hfill
    \begin{subfigure}{0.45\textwidth}
      \centering
        \includegraphics[width=\linewidth]{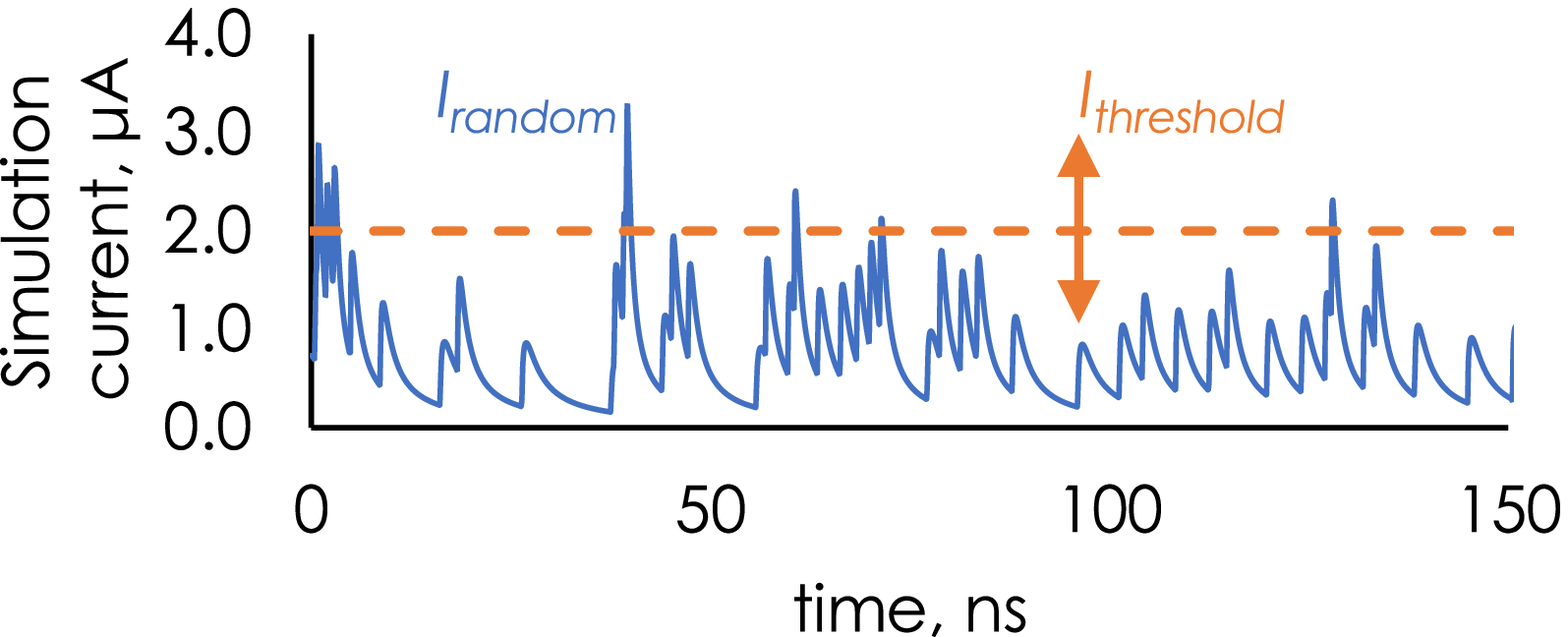}
        \caption{Simulated random current}
        \label{subfig:rand_current}
    \end{subfigure}
    
    \vspace{0.1in}
    
    \begin{subfigure}[b]{0.24\textwidth}
        \centering
        \includegraphics[width=1.5in]{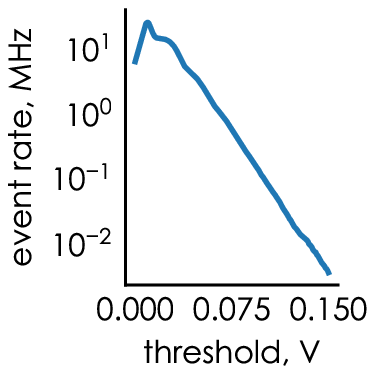}
        \caption{Meas. transfer function}
        \label{subfig:measured_xfer_fn}
    \end{subfigure}
    \begin{subfigure}[b]{0.24\textwidth}
        \centering
        \includegraphics[width=1.5in]{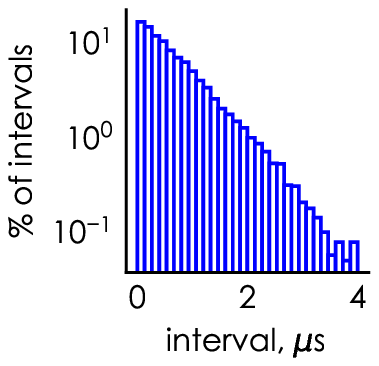}
        \caption{Meas. timing statistics}
        \label{subfig:measured_statistics}
    \end{subfigure}
    \begin{subfigure}[b]{0.24\textwidth}
        \centering
        \includegraphics[width=1.5in]{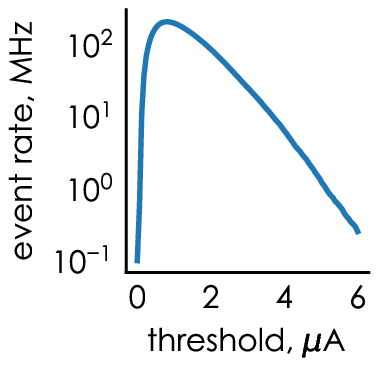}
        \caption{Sim. transfer function}
        \label{subfig:sim_xfer_fn}
    \end{subfigure}
    \begin{subfigure}[b]{0.24\textwidth}
        \centering
        \includegraphics[width=1.5in]{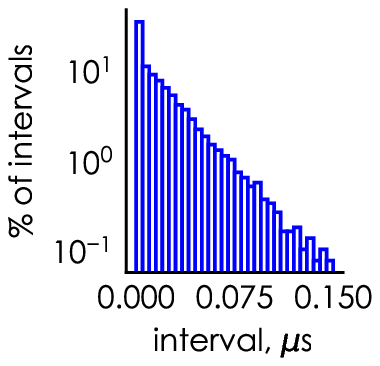}
        \caption{Sim. timing statistics}
        \label{subfig:sim_statistics}
    \end{subfigure}

\end{center}
\caption 
{ \label{fig:variable_rate_spads}
Design of variable-rate SPAD circuits.  Two designs are shown, an experimental design (a) made of discrete components for which we have measured results and a simulated design (b) meant to establish that the variable-rate behavior is not dependent on the specific characteristics of SiPMs.  For both designs, the pulse rate can be controlled with an exponential transfer function (e, g) in accordance to equation \ref{eqn:spad_requirement} and the interval between pulses is shown to be an exponential (f, h), a defining feature of Poisson processes. } 
\end{figure} 

\subsection{Variable-rate SPAD circuits}

SPADs have two physical controls for their detection rate, and further control over their effective detection rate can be created by filtering circuits.  The first physical control is the illumination incident upon the SPAD, which provides a large dynamic range for how often the SPAD triggers; the second is the reverse bias of the SPAD, which can tune the dark count rate and the photon detection efficiency \cite{commericial_spad_characterization}.  Since illumination can easily be controlled across an entire chip but not locally, it is suitable for temperature control but not for controlling the rate of individual SPADs. Controlling the reverse bias can be done for each SPAD individually, but the effect of changing the SPAD bias is dependent on the exact SPAD design; some SPAD designs demonstrate bias effects exponentially altering the dark count rate \cite{exponential_dcr}, while others do not \cite{28nm_spad}.  For these reasons a filtering and comparison circuit, shown in figure \ref{fig:variable_rate_spads}, is used for the current design instead.  By filtering the output of a SPAD on a time scale close to the SPAD's avalanche rate, the filtered output $V_{random}$ or $I_{random}$ continues to pulse at each photon detection but with a magnitude that is now variable, depending on the history of SPAD avalanches.  The variable-height pulses can then be compared to a threshold, yielding a decreasing \textit{detected} pulse rate as the threshold is increased.  Whether or not the resulting detected pulse rate is suitable for construction of a CTMC neuron depends on the distribution of the random filtered signal, which in general is hard to predict exactly, especially if there is nonlinear filtering.  One tractable scenario is the limit as the filtering timescale $ {\tau}_f $ becomes much longer than the photon arrival timescale $ {\tau}_p $.  In this case the sum of pulses may be simply treated as an infinite sum of random variables, and the central limit theorem indicates the filtered signal will be Gaussian.  Since thresholding integrates the tail of the filtered distribution, the threshold to pulse rate transfer function will be proportional to the complementary error function, $r \propto erfc(E/{T'})$.  While not the desired transfer function, it may still be used as a rough approximation to an exponential transfer function.  Fortunately, experimental and simulated results show  when ${\tau}_p \approx {\tau}_f$ this filtering concept can yield the exponential transfer function required by equation \ref{eqn:spad_requirement}, and the interval between detected pulses is still nearly a Poisson process as required by the CTMC analysis.

We simulated a CMOS version of the filtered SPAD design  based on a model of SPADs in a CMOS process \cite{SPAD_model_in_use} .  The SPAD model produces avalanches of identical magnitude which when processed by a Gm-C filter, as shown in figure \ref{subfig:sim_circuit}, produces an exponential distribution of current pulses.  A sample of the filtered pulses is shown in figure \ref{subfig:rand_current}, and the threshold-rate transfer function is shown in figure \ref{subfig:sim_xfer_fn}.  This design is provided as an example, and there are undoubtedly other filtering methods capable of yielding an exponential distribution of pulse heights.

\begin{figure}
\begin{center}

    \hfill
    \begin{subfigure}[b]{0.55\textwidth}
        \centering
        \includegraphics[width=0.9\textwidth]{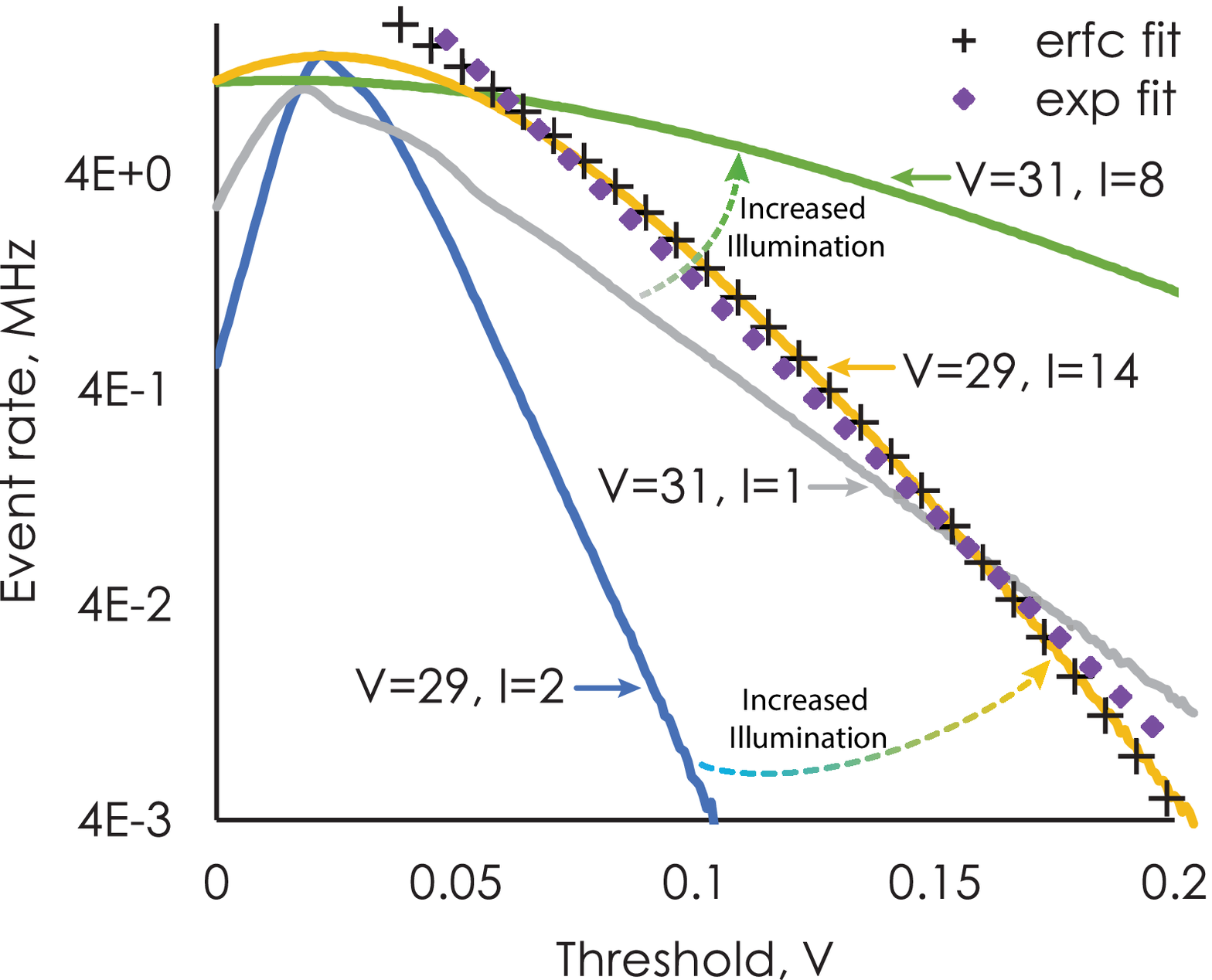}
        \caption{Different operating conditions}
        \label{fig:bias_illumination_conditions}
    \end{subfigure}
    \hfill
    \begin{subfigure}[b]{0.4\textwidth}
        \centering
        \includegraphics[width=\textwidth]{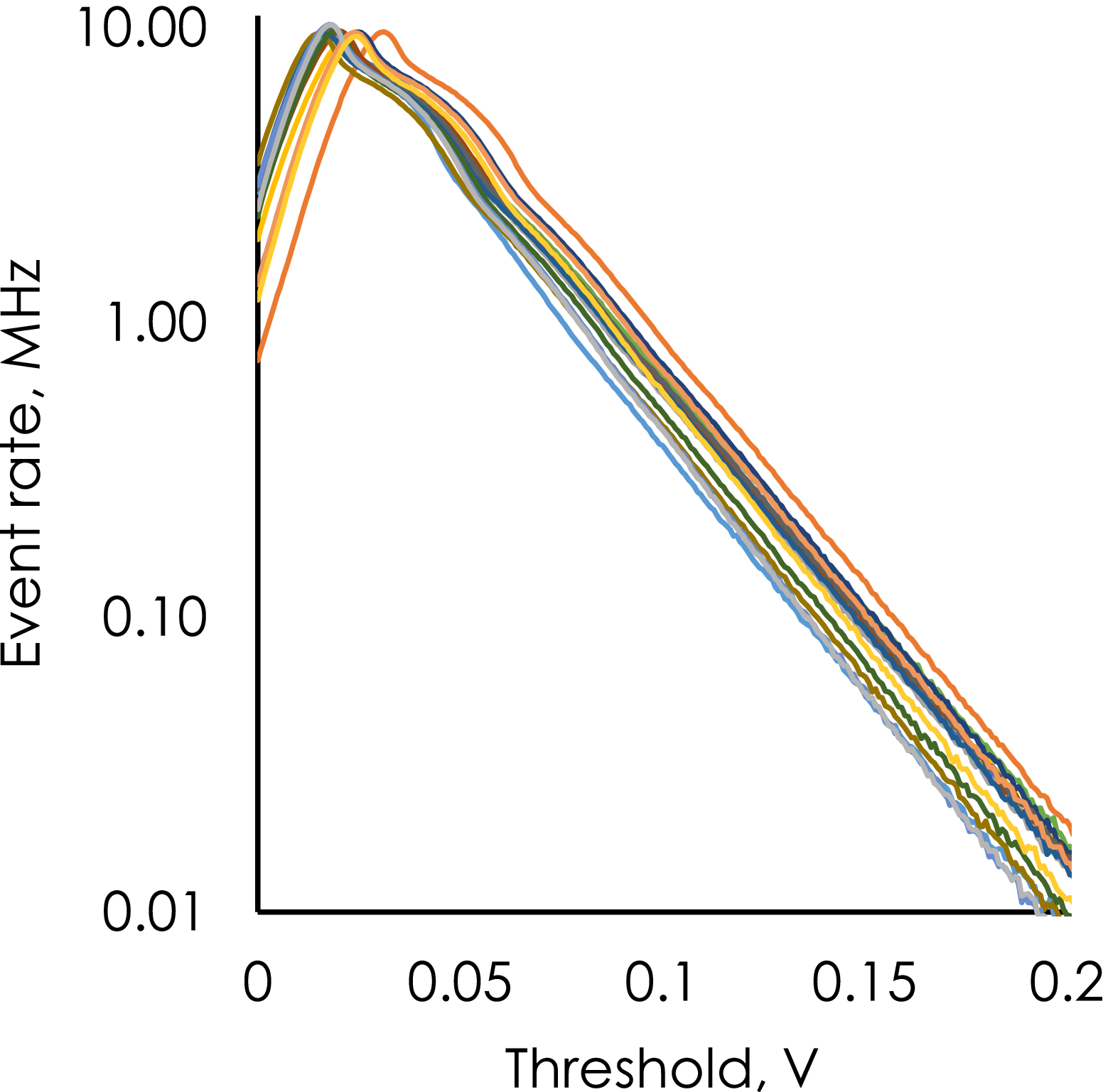}
        \caption{SPAD variability}
        \label{fig:spad_variability}
    \end{subfigure}
    \hfill

\end{center}
\caption 
{ \label{fig:behavior_of_experimental_spad_circuits}
Control and uniformity of variable-rate SPAD circuit transfer functions. (a) The slope of the transfer function can be controlled by the bias voltage $V$ and the illumination $I$.  At high illumination, the transfer functions fit the complimentary error function better than an exponential. (b) The transfer functions of the 16 variable rate SPADs in the test setup, showing threshold variation but identical slopes.} 
\end{figure}

Our experimental variable rate event generators operate on another implementation of the filtering principle.  Due to sourcing constraints, we physically realized variable rate event generation using silicon photomultipliers (SiPMs) rather than SPADs.  As an array of SPADs, SiPMs perform the same key function but have additional behaviors, such as some built-in filtering and the possibility of simultaneous avalanches.  Despite these differences, a SiPM is not fundamentally different from a single SPAD if filtering is involved.  Our experimental SiPM circuit, including internal SiPM filtering elements, is shown in figure \ref{subfig:exp_circuit}.  This circuit was found to have a sufficiently exponential transfer function under a range of SPAD bias and illumination conditions, with a few cases shown in figure \ref{fig:bias_illumination_conditions}.  At high illumination the transfer function looks less like an exponential and more like the tail of a complementary error function, as expected when the SPAD avalanche rate becomes much faster than the filtering time constant; nonetheless, the exponential is still a close fit and under high illumination the Ising and Potts models still produce accurate statistics.  Another key feature visible in figure \ref{fig:bias_illumination_conditions} is the variation in the transfer function slope as the illumination and bias are changed, which allows the computational temperature of an Ising or Potts model to be easily controlled in real time. Unlike many current designs  the model weights do not need to be reconfigured.  While the transfer function slopes of each SPAD can be easily controlled, they are also quite stable between SPAD circuits under the same operating conditions.  Figure \ref{fig:spad_variability} shows a simple offset term can match the 16 different experimental variable-rate SPAD circuits. 

\subsection{Experimental Demonstrations}
\begin{figure}[!t]
\begin{center}

    \begin{subfigure}[b]{0.49\textwidth}
        \centering
        \includegraphics[width=2.9in]{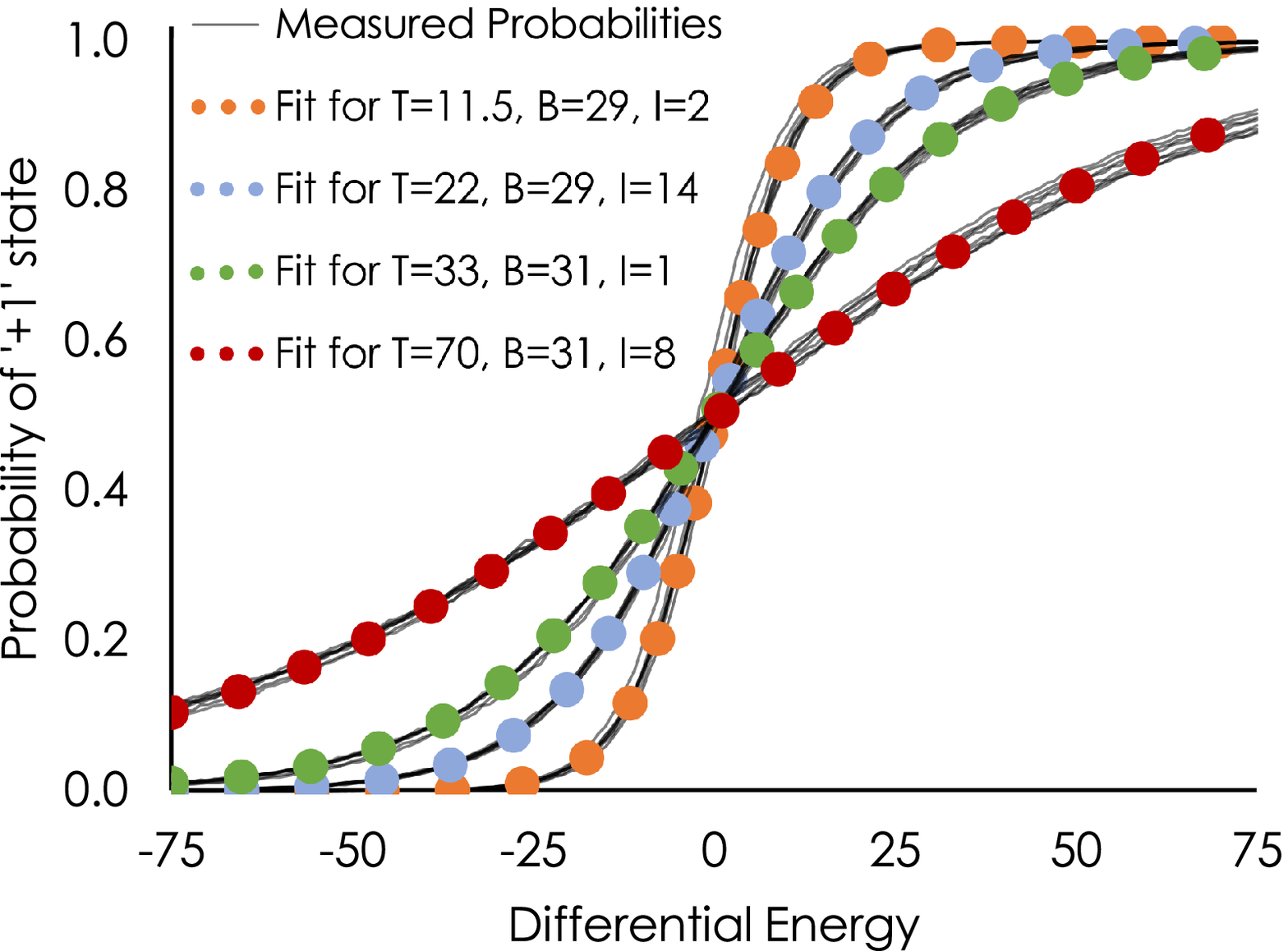}
        \caption{Ising neuron probabilities}
        \label{fig:tanh}
    \end{subfigure}
    \hfill
    \begin{subfigure}[b]{0.49\textwidth}
        \centering
        \includegraphics[width=2.9in]{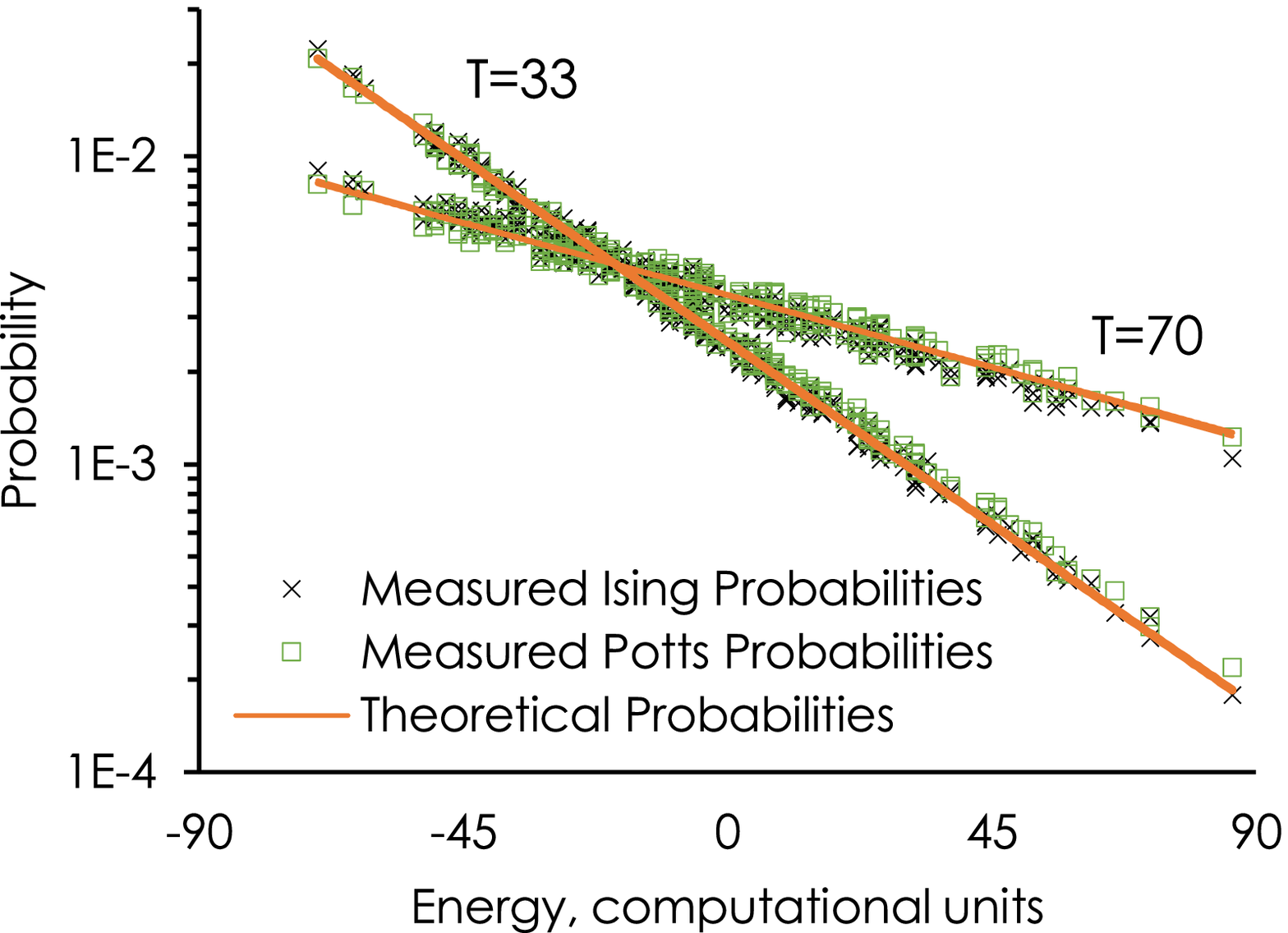}
        \caption{Probabilities of a model with random weights}
        \label{fig:256_probabilities}
    \end{subfigure}
    \hfill
    
    \vspace{0.2in}
    
    \hfill
    \begin{subfigure}[b]{4in}
        \centering
        \includegraphics[width=4in]{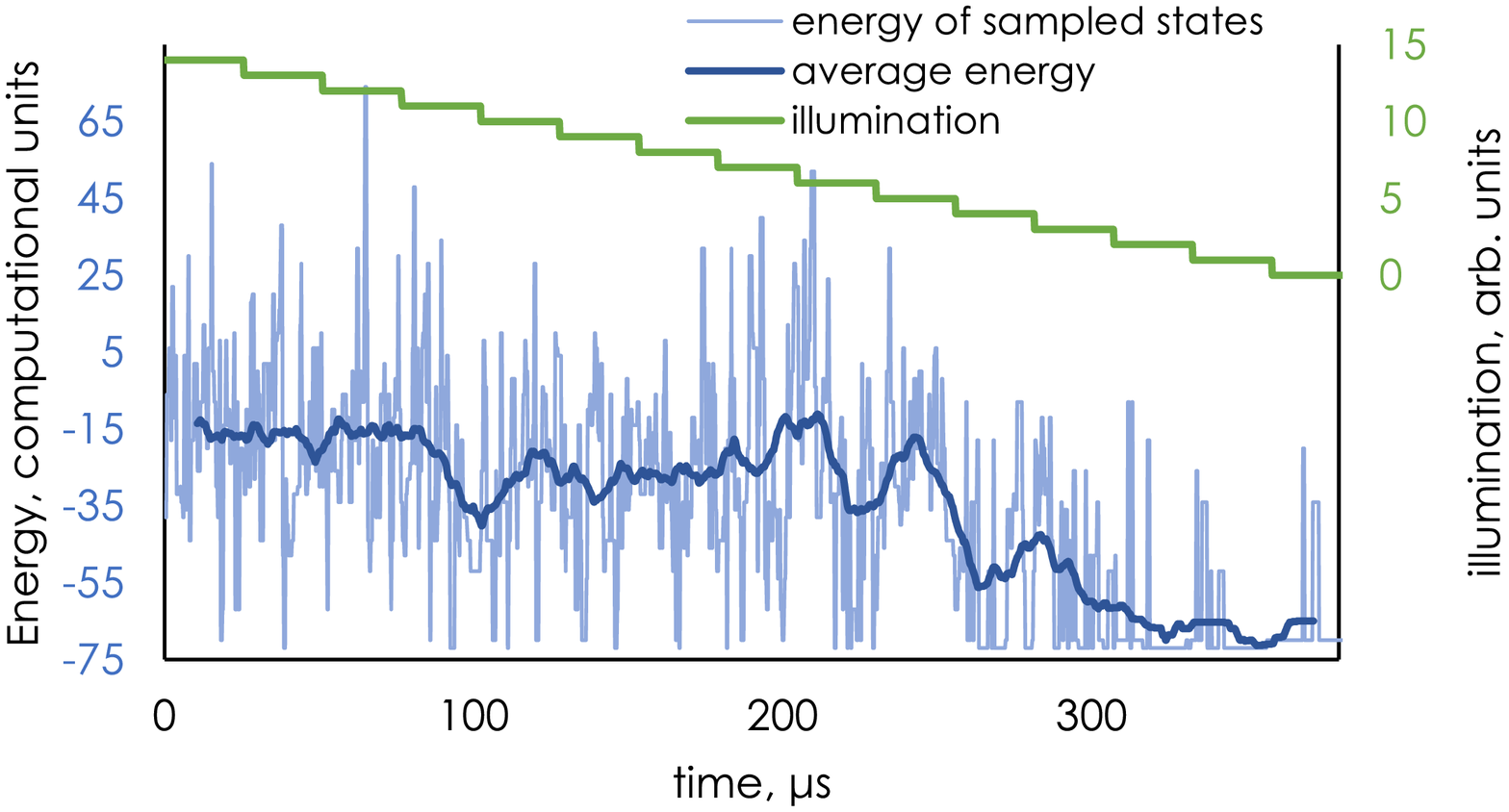}
        \caption{Annealing in the Ising model}
        \label{fig:Ising_annealing}
    \end{subfigure}
    \hfill
    \begin{subfigure}[b]{1.7in}
        \centering
        \raisebox{0.2in}{\includegraphics[width=1.7in]{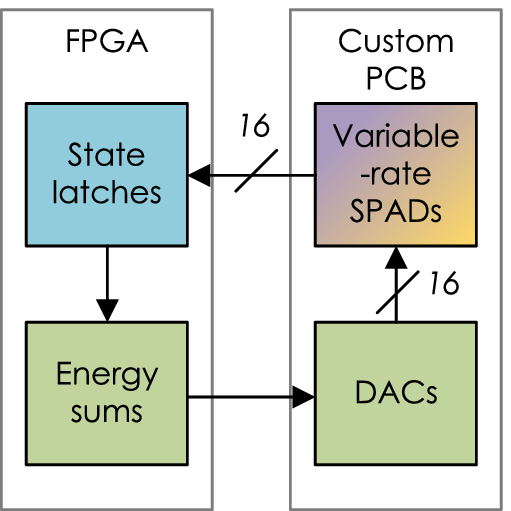}}
        \caption{Implementation}
        \label{fig:lab_diagram}
    \end{subfigure}
    \hfill
    \hfill
    
\end{center}
\caption 
{ \label{fig:experimental_system}
Experimental methods (d) and measured results.  SPAD CTMC neurons demonstrate highly accurate statistical behavior, both at the level of individual Ising neurons (a) and when used to form complete Ising and Potts Boltzmann machines (b).  (c) Annealing can be easily controlled by adjusting the illumination on the SPADs.} 
\end{figure}

The first step in assembling the Ising and Potts models is calibration of individual neurons, which in the Ising case can be verified by comparing the transfer function of each neuron to equation \ref{eqn:tanh}.  Calibration serves to compensate for differences in slope and offset in the exponential transfer functions of each variable-rate SPAD, although for our experimental variable rate event generators, slope compensation was not needed since the slopes of the transfer functions in figure \ref{fig:spad_variability} are all nearly the same.  Calibration is performed for each SPAD circuit individually.  A target event rate is picked, and for each SPAD the energy bias yielding that event rate is determined to be the zero operating point for that SPAD circuit.  When assembled into Ising neurons, the calibration and the performance of the CTMC neuron as a whole is verified by measuring the neuron's transfer function. The differential energy sum is treated as an input, and the probability with which the neuron is in its $+1$ state is treated as an output.  Most proposed stochastic bit designs are evaluated this way \cite{MTJ_for_pbits,memristor_Ising_machine}.  Since the output of the neuron transfer function is a probability, it is measured by averaging many states as the input energy is held constant, and the input is then changed and the procedure is repeated.  Figure \ref{fig:tanh} shows the measured transfer function for eight Ising nodes, constructed out of an array of 16 variable rate SiPM circuits.  Transfer functions are shown for four combinations of illumination and SPAD bias, which manifests as different computational temperatures (tanh slopes). While the variable rate SPAD circuit produces a complementary error function transfer function under high illumination rather than an exponential, the Ising model transfer function is still accurate showing the high resilience of the CTMC architecture to imperfections.  Fits of equation \ref{eqn:tanh} are included for each illumination condition, which also yields the computational temperature which we later use to generate matching theoretical distributions for evaluation of full Ising and Potts models.  Unfortunately no similar, simple verification applies to Potts model neurons, but the quality of both Ising and Potts neurons is demonstrated indirectly by the performance of the complete models.

%With the performance of individual neurons verified, the next step is to demonstrate correct behavior when multiple Ising and Potts neurons are interacting with each other.  
The complete Ising and Potts machines comprise an FPGA portion and a custom PCB portion, shown in figure \ref{fig:lab_diagram}.  This arrangement is similar to that used in other demonstrations \cite{MTJ_for_pbits, memristor_Ising_machine}.  The FPGA takes SPAD pulses as inputs, implements the latching circuits, digitally calculates energy summations, and writes the energy summations back to the variable rate SPAD circuits as an analog value using digital to analog converters.  The custom PCB has a 4x4 array of SiPMs which are coupled to 16 comparators, yielding 16 variable-rate SPAD circuits.  Since latching is performed in the FPGA and is programmable, we use the 16 variable-rate SPAD circuits to demonstrate fully-connected networks of either eight Ising nodes or four $q=4$ Potts nodes. Other configurations are also possible using the same hardware.  Since the number of states is small, the statistical correctness of the models can be directly established.  We chose to work with arbitrary models with random weights, which is the most straightforward way to use all available neurons and weights.  In this case the interaction between neurons is maximized so that any issues in statistical sampling are more likely to appear. Furthermore, since the number of states $2^8=4^4=256$ in the Ising demonstration and the Potts demonstration are the same, the Ising and Potts models can be configured to sample from the same distribution which allows comparison to each other as well as to the ideal theoretical Boltzmann distribution.  First weights and biases for the Ising model are selected uniformly from the integers on $[-8, 8]$, and then the Ising model weights are converted to equivalent Potts model weights (see supplementary material).  For a given set of SPAD illumination and bias conditions, the theoretical Boltzmann distribution is calculated from the Ising network weights and the computational temperature found in figure \ref{fig:tanh}.   Five million samples each are measured from the Ising and Potts machines to determine the experimental distributions.  Figure \ref{fig:256_probabilities} compares calculated, Ising, and Potts distributions at two different temperatures, all with the same set of random weights.  To illustrate that the Ising and Potts models sample from a Boltzmann distribution, each of the 256 states are plotted according to their energy and measured probability, fitting to a straight line on a logarithmic plot.  In this format it is easy to tell that the SPAD CTMC Ising and Potts models match the desired distribution, achieving a Kullback-Leibler divergence of 0.01.  While states are not sampled exactly as often as they should be, any significant difference in energy between states is reflected in their probability and the lowest energy states are universally the most probable.  This statistical correctness establishes that SPAD CTMC neurons will be able to find the lowest energy configuration of any real-world problem.

The computational temperature of SPAD CTMC neurons can be controlled continuously using several methods, including changing the illumination on the SPADs.  Normally, a gradual decrease in computational temperature allows an annealing processor to gradually settle into a single low energy state.   Our SiPM CTMC neurons allow a factor of two temperature change in response to illumination, however since the demonstration network is limited to eight Ising neurons or four Potts neurons, the model does not lock into a single state at the lower temperature, and the lowest energy state can easily be reached even during sampling at higher temperatures.  Nonetheless, by gradually decreasing the illumination and thus the computational temperature, the average energy that the model samples from can be seen to decrease.  Figure \ref{fig:Ising_annealing} shows the energy and time-averaged energy of the sampled states during the annealing schedule defined by the illumination, demonstrating the inherent annealing capability of SPAD CTMC neurons.  In contrast to the distribution measurements taken at different temperatures, during annealing the CTMC neurons can only be calibrated once for a single temperature; fortunately the calibration is roughly applicable across temperatures.  However, there are some limitations, such as an inability to approach zero computational temperature.  In addition it can be difficult to independently control the temperature and the mean event rate of the SPADs.  However, as an analog neuron, the temperature of the SPAD CTMC neuron may be controlled by other methods as well such as a global analog weight scaling factor.  Regardless of the exact design, there are several simple temperature control options when using SPAD CTMC neurons that can be applied continuously without pausing the operation of the annealing processor.

\section{Discussion}
A proof-of-concept for using single-photon avalanche diodes (SPADs) as the main computational element in annealing processors has been presented.  Since SPADs produce random events rather than random states, they must be coupled to latching circuitry to form Ising and Potts nodes.  The resulting circuits form stochastic finite state machines whose state is a continuous-time Markov chain (CTMC).  While this is schematically more complex than other proposed designs for hardware Ising nodes, it allows for much better control over the behavior of each node and is the key enabler of Potts model annealing in addition to the usual Ising model annealing. SPAD CTMC neurons are additionally compatible with CMOS fabrication processes and offer many options for temperature control. CMOS integration requires a few considerations since SPAD designs have been mainly optimized for imaging applications which have different constraints compared to annealing processors. A prime example is the dark count rate which needs to be relatively low for imagers but not for annealing processors, relaxing some of the design constraints for CTMC SPAD neurons and improving scaling\cite{SPAD_scaling}.
One area of future work is re-imagining the Annealing processor architecture to make effective use of the Potts model, since the weight configuration options for the Potts model are far greater than for the Ising model.  When these issues are addressed, annealing processors based on SPAD CTMC neurons will be able to solve more complex optimization problems than current annealing processors do, via both scaling to a greater number of nodes and the step improvement in representational ability bestowed by the Potts model.

\bibliography{references}

\begin{thebibliography}{10}
\providecommand{\url}[1]{{#1}}
\providecommand{\urlprefix}{URL }
\providecommand{\doi}[1]{\url{https://doi.org/#1}}
\bibcommenthead

\bibitem{original_simulated_annealing}
S.~Kirkpatrick, C.D. Gelatt, M.P. Vecchi, Optimization by simulated annealing.
\newblock Science \textbf{220}(4598), 671--680 (1983).
\newblock \doi{10.1126/science.220.4598.671}.
\newblock \urlprefix\url{https://doi.org/10.1126/science.220.4598.671}

\bibitem{simulated_annealing_chip_floorplanning}
R.~Rutenbar, Simulated annealing algorithms: an overview.
\newblock IEEE Circuits and Devices Magazine \textbf{5}(1), 19--26 (1989).
\newblock \doi{10.1109/101.17235}

\bibitem{simulated_annealing_in_operations_research}
C.~Koulamas, S.~Antony, R.~Jaen, A survey of simulated annealing applications
  to operations research problems.
\newblock Omega \textbf{22}(1), 41--56 (1994).
\newblock \doi{https://doi.org/10.1016/0305-0483(94)90006-X}.
\newblock
  \urlprefix\url{https://www.sciencedirect.com/science/article/pii/030504839490006X}

\bibitem{Ising_formulations}
A.~Lucas, Ising formulations of many np problems.
\newblock Frontiers in Physics \textbf{2} (2014).
\newblock \doi{10.3389/fphy.2014.00005}.
\newblock
  \urlprefix\url{https://www.frontiersin.org/article/10.3389/fphy.2014.00005}

\bibitem{review_of_hw_Ising_solvers}
N.~Mohseni, P.L. McMahon, T.~Byrnes, Ising machines as hardware solvers of
  combinatorial optimization problems.
\newblock Nature Reviews Physics \textbf{4}(6), 363--379 (2022).
\newblock \doi{10.1038/s42254-022-00440-8}.
\newblock \urlprefix\url{https://doi.org/10.1038/s42254-022-00440-8}

\bibitem{MTJ_for_pbits}
K.Y. Camsari, S.~Salahuddin, S.~Datta, Implementing p-bits with embedded mtj.
\newblock IEEE Electron Device Letters \textbf{38}(12), 1767--1770 (2017).
\newblock \doi{10.1109/LED.2017.2768321}

\bibitem{2x30k_CMOS_annealer}
T.~Takemoto, M.~Hayashi, C.~Yoshimura, M.~Yamaoka, A 2 $\times$ 30k-spin
  multi-chip scalable cmos annealing processor based on a processing-in-memory
  approach for solving large-scale combinatorial optimization problems.
\newblock IEEE Journal of Solid-State Circuits \textbf{55}(1), 145--156 (2020).
\newblock \doi{10.1109/JSSC.2019.2949230}

\bibitem{compute_in_memory_CMOS_annealer}
Y.~Su, H.~Kim, B.~Kim, Cim-spin: A scalable cmos annealing processor with
  digital in-memory spin operators and register spins for combinatorial
  optimization problems.
\newblock IEEE Journal of Solid-State Circuits \textbf{57}(7), 2263 -- 2273
  (2022).
\newblock \doi{10.1109/JSSC.2021.3139901}

\bibitem{massively_parallel_sparse_Ising}
N.A. Aadit, A.~Grimaldi, M.~Carpentieri, L.~Theogarajan, J.M. Martinis,
  G.~Finocchio, K.Y. Camsari, Massively parallel probabilistic computing with
  sparse ising machines.
\newblock Nature Electronics  (2022).
\newblock \doi{10.1038/s41928-022-00774-2}.
\newblock \urlprefix\url{https://doi.org/10.1038/s41928-022-00774-2}

\bibitem{series_parallel_RBM_in_FPGA}
K.~Ueyoshi, T.~Marukame, T.~Asai, M.~Motomura, A.~Schmid, {FPGA} implementation
  of a scalable and highly parallel architecture for restricted boltzmann
  machines.
\newblock Circuits and Systems \textbf{07}(09), 2132--2141 (2016).
\newblock \doi{10.4236/cs.2016.79185}.
\newblock \urlprefix\url{https://doi.org/10.4236/cs.2016.79185}

\bibitem{DEC_coprocessor_BM}
M.~Skubiszewski, in \emph{Proceedings of the Fourth IEEE Symposium on Parallel
  and Distributed Processing} (1992), pp. 107--110

\bibitem{memristor_Ising_machine}
X.~Yan, J.~Ma, T.~Wu, A.~Zhang, J.~Wu, M.~Chin, Z.~Zhang, M.~Dubey, W.~Wu,
  M.S.W. Chen, J.~Guo, H.~Wang, Reconfigurable stochastic neurons based on tin
  {oxide/MoS2} hetero-memristors for simulated annealing and the boltzmann
  machine.
\newblock Nature Communications \textbf{12}(1), 5710 (2021)

\bibitem{memristor_weight_noise_hopfield_Ising}
F.~Cai, S.~Kumar, T.~Van~Vaerenbergh, X.~Sheng, R.~Liu, C.~Li, Z.~Liu,
  M.~Foltin, S.~Yu, Q.~Xia, J.J. Yang, R.~Beausoleil, W.D. Lu, J.P. Strachan,
  Power-efficient combinatorial optimization using intrinsic noise in memristor
  hopfield neural networks.
\newblock Nature Electronics \textbf{3}(7), 409--418 (2020)

\bibitem{coherent_Ising_perspective}
Y.~Yamamoto, T.~Leleu, S.~Ganguli, H.~Mabuchi, Coherent ising
  machines---quantum optics and neural network perspectives.
\newblock Applied Physics Letters \textbf{117}(16), 160,501 (2020)

\bibitem{atomic_boltzmann}
B.~Kiraly, E.J. Knol, W.M.J. van Weerdenburg, H.J. Kappen, A.A. Khajetoorians,
  An atomic boltzmann machine capable of self-adaption.
\newblock Nature Nanotechnology \textbf{16}(4), 414--420 (2021).
\newblock \doi{10.1038/s41565-020-00838-4}.
\newblock \urlprefix\url{https://doi.org/10.1038/s41565-020-00838-4}

\bibitem{efficient_tanh_example}
B.~Zamanlooy, M.~Mirhassani, Efficient vlsi implementation of neural networks
  with hyperbolic tangent activation function.
\newblock IEEE Transactions on Very Large Scale Integration (VLSI) Systems
  \textbf{22}(1), 39--48 (2014).
\newblock \doi{10.1109/TVLSI.2012.2232321}

\bibitem{analog_bistable_Ising_machine}
R.~Afoakwa, Y.~Zhang, U.K.R. Vengalam, Z.~Ignjatovic, M.~Huang, in \emph{2021
  IEEE International Symposium on High-Performance Computer Architecture
  (HPCA)} (2021), pp. 749--760.
\newblock \doi{10.1109/HPCA51647.2021.00068}

\bibitem{analog_LC_oscillator_Ising}
J.~Chou, S.~Bramhavar, S.~Ghosh, W.~Herzog, Analog coupled oscillator based
  weighted ising machine.
\newblock Scientific Reports \textbf{9}(1) (2019).
\newblock \doi{10.1038/s41598-019-49699-5}.
\newblock \urlprefix\url{https://doi.org/10.1038/s41598-019-49699-5}

\bibitem{ising_history}
S.G. Brush, History of the {Lenz-Ising} model.
\newblock Reviews of Modern Physics \textbf{39}(4), 883--893 (1967)

\bibitem{potts_history}
F.Y. Wu, The potts model.
\newblock Reviews of Modern Physics \textbf{54}(1), 235--268 (1982)

\bibitem{potts_superiority}
C.~Peterson, B.~S\"{o}derberg, A new method for mapping optimization problems
  onto neural networks.
\newblock International Journal of Neural Systems \textbf{01}(01), 3--22
  (1989).
\newblock \doi{10.1142/s0129065789000414}.
\newblock \urlprefix\url{https://doi.org/10.1142/s0129065789000414}

\bibitem{potts_vs_Ising_natural_representation}
B.~S{\"o}derberg, in \emph{Scientific Applications of Neural Nets}, ed. by J.W.
  Clark, T.~Lindenau, M.L. Ristig (Springer Berlin Heidelberg, Berlin,
  Heidelberg, 1999), pp. 243--256

\bibitem{binary_vs_onehot_encoding}
S.~Okada, M.~Ohzeki, S.~Taguchi, Efficient partition of integer optimization
  problems with one-hot encoding.
\newblock Scientific Reports \textbf{9}(1), 13,036 (2019)

\bibitem{potts_graph_coloring}
I.~Kanter, H.~Sompolinsky, Graph optimisation problems and the potts glass.
\newblock Journal of Physics A: Mathematical and General \textbf{20}(11),
  L673--L679 (1987).
\newblock \doi{10.1088/0305-4470/20/11/001}.
\newblock \urlprefix\url{https://doi.org/10.1088/0305-4470/20/11/001}

\bibitem{simulated_annealing_book}
E.~Aarts, J.~Korst, \emph{Simulated annealing and Boltzmann machines}.
\newblock Wiley Series in Discrete Mathematics \& Optimization (John Wiley \&
  Sons, Chichester, England, 1988)

\bibitem{original_gibbs_sampler}
S.~Geman, D.~Geman, Stochastic relaxation, gibbs distributions, and the
  bayesian restoration of images.
\newblock IEEE Transactions on Pattern Analysis and Machine Intelligence
  \textbf{PAMI-6}(6), 721--741 (1984).
\newblock \doi{10.1109/TPAMI.1984.4767596}

\bibitem{spad_qrng_1}
A.~Stanco, D.G. Marangon, G.~Vallone, S.~Burri, E.~Charbon, P.~Villoresi,
  Efficient random number generation techniques for cmos single-photon
  avalanche diode array exploiting fast time tagging units.
\newblock Physics Review Research \textbf{2}, 023,287 (2020).
\newblock \doi{10.1103/PhysRevResearch.2.023287}.
\newblock
  \urlprefix\url{https://link.aps.org/doi/10.1103/PhysRevResearch.2.023287}

\bibitem{spad_qrng_2}
A.~Tontini, L.~Gasparini, N.~Massari, R.~Passerone, Spad-based quantum random
  number generator with an $n^{\rm{th}}$ -order rank algorithm on fpga.
\newblock IEEE Transactions on Circuits and Systems II: Express Briefs
  \textbf{66}(12), 2067--2071 (2019).
\newblock \doi{10.1109/TCSII.2019.2909013}

\bibitem{1um_130nm_spad}
Z.~You, L.~Parmesan, S.~Pellegrini, R.K. Henderson, in \emph{International
  Image Sensor Workshop} (2017)

\bibitem{spad_imager_review}
C.~Bruschini, H.~Homulle, I.M. Antolovic, S.~Burri, E.~Charbon, Single-photon
  avalanche diode imagers in biophotonics: review and outlook.
\newblock Light: Science and Applications \textbf{8}(1), 87 (2019)

\bibitem{hybrid_potts_on_CIM}
K.~Inaba, T.~Inagaki, K.~Igarashi, S.~Utsunomiya, T.~Honjo, T.~Ikuta,
  K.~Enbutsu, T.~Umeki, R.~Kasahara, K.~Inoue, Y.~Yamamoto, H.~Takesue, Potts
  model solver based on hybrid physical and digital architecture.
\newblock Communications Physics \textbf{5}(1) (2022).
\newblock \doi{10.1038/s42005-022-00908-0}.
\newblock \urlprefix\url{https://doi.org/10.1038/s42005-022-00908-0}

\bibitem{Potts_on_Ising_via_halfhot}
S.~Okada, M.~Ohzeki, K.~Tanaka, Efficient quantum and simulated annealing of
  potts models using a half-hot constraint.
\newblock Journal of the Physical Society of Japan \textbf{89}(9), 094,801
  (2020).
\newblock \doi{10.7566/jpsj.89.094801}.
\newblock \urlprefix\url{https://doi.org/10.7566/jpsj.89.094801}

\bibitem{bose-einstein_condensate_potts}
K.P. Kalinin, N.G. Berloff, Simulating ising and $n$-state planar potts models
  and external fields with nonequilibrium condensates.
\newblock Physics Review Letters \textbf{121}, 235,302 (2018).
\newblock \doi{10.1103/PhysRevLett.121.235302}.
\newblock
  \urlprefix\url{https://link.aps.org/doi/10.1103/PhysRevLett.121.235302}

\bibitem{Hopfield_digital_weights_current_sum}
P.~Hollis, J.~Paulos, Artificial neural networks using mos analog multipliers.
\newblock IEEE Journal of Solid-State Circuits \textbf{25}(3), 849--855 (1990).
\newblock \doi{10.1109/4.102684}

\bibitem{CCT_CMOS_analog_synapse}
X.~Gu, Z.~Wan, S.S. Iyer, Charge-trap transistors for cmos-only analog memory.
\newblock IEEE Transactions on Electron Devices \textbf{66}(10), 4183--4187
  (2019).
\newblock \doi{10.1109/TED.2019.2933484}

\bibitem{analog_winner_take_all}
J.~Lazzaro, S.~Ryckebusch, M.~Mahowald, C.A. Mead, in \emph{Advances in Neural
  Information Processing Systems}, vol.~1, ed. by D.~Touretzky
  (Morgan-Kaufmann, 1988).
\newblock
  \urlprefix\url{https://proceedings.neurips.cc/paper/1988/file/a8f15eda80c50adb0e71943adc8015cf-Paper.pdf}

\bibitem{commericial_spad_characterization}
M.~Stipčević, D.~Wang, R.~Ursin, Characterization of a commercially available
  large area, high detection efficiency single-photon avalanche diode.
\newblock Journal of Lightwave Technology \textbf{31}(23), 3591--3596 (2013).
\newblock \doi{10.1109/JLT.2013.2286422}

\bibitem{exponential_dcr}
X.~Lu, M.K. Law, Y.~Jiang, X.~Zhao, P.I. Mak, R.P. Martins, A 4-um diameter
  spad using less-doped n-well guard ring in baseline 65-nm cmos.
\newblock IEEE Transactions on Electron Devices \textbf{67}(5), 2223--2225
  (2020).
\newblock \doi{10.1109/TED.2020.2982701}

\bibitem{28nm_spad}
T.C. de~Albuquerque, F.~Calmon, R.~Clerc, P.~Pittet, Y.~Benhammou, D.~Golanski,
  S.~Jouan, D.~Rideau, A.~Cathelin, in \emph{2018 48th European Solid-State
  Device Research Conference (ESSDERC)} (2018), pp. 82--85.
\newblock \doi{10.1109/ESSDERC.2018.8486852}

\bibitem{SPAD_model_in_use}
J.M. L{\'{o}}pez-Mart{\'{i}}nez, I.~Vornicu, R.~Carmona-Gal{\'{a}}n,
  {\'{A}}.~Rodríguez-V{\'{a}}zquez, in \emph{2018 25th IEEE International
  Conference on Electronics, Circuits and Systems (ICECS)} (2018), pp.
  137--140.
\newblock \doi{10.1109/ICECS.2018.8617962}

\bibitem{SPAD_scaling}
K.~Morimoto, E.~Charbon, A scaling law for spad pixel miniaturization.
\newblock Sensors \textbf{21}(10) (2021).
\newblock \doi{10.3390/s21103447}.
\newblock \urlprefix\url{https://www.mdpi.com/1424-8220/21/10/3447}

\end{thebibliography}

\end{document}

\maketitle

\renewcommand{\thetable}{S\arabic{table}}%adds S before table and fig numbers
\renewcommand{\thefigure}{S\arabic{figure}}%

\pagebreak
\section{Scalable one-hot latching circuit}

\begin{figure}[!t]
\begin{center}
    \begin{subfigure}[b]{6.5in}
        \centering
        \includegraphics[width=3in]{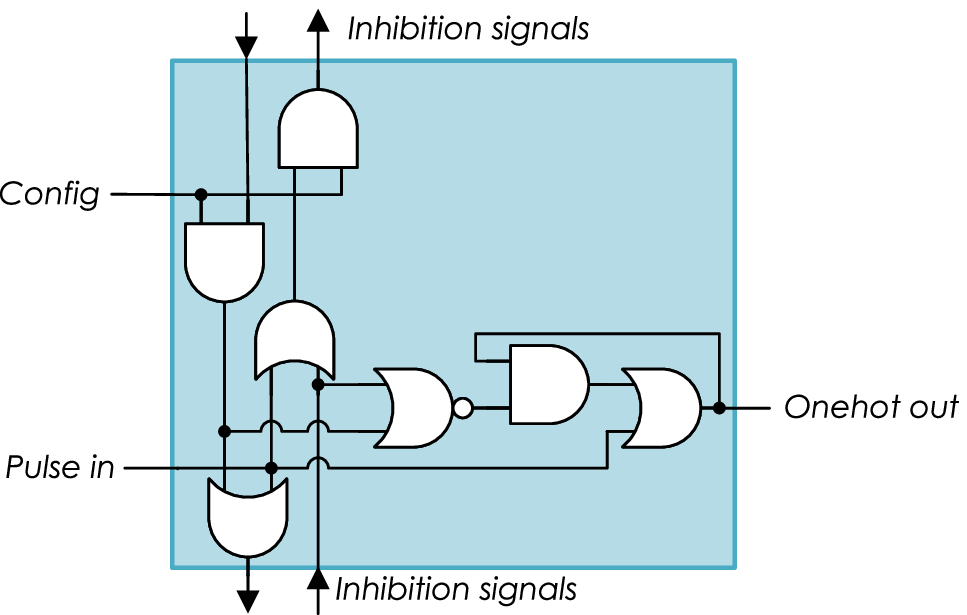}
    \end{subfigure}
    \begin{subfigure}[b]{6.5in}
        \centering
        \includegraphics[width=3in]{scalable_onehot_latch.eps}
    \end{subfigure}
\end{center}
\caption 
{ \label{fig:onehot_circuit}
Logic design of a scalable one-hot latch.} 
\end{figure}

The one-hot latch required for Potts model CTMC neurons can be implemented as in figure \ref{fig:onehot_circuit}.  This circuit is composed of a chain of repeating elements (blue rectangle; two are shown).  Each circuit has a pulse input that when high, sends an 'inhibition' signal cascading through all connected elements which sets the output of all elements to zero.  Meanwhile, the output of the latching element with the pulse input is set to high (and sees no inhibition signal); thus for $q$ connected elements, one will always have a high output and the rest will have low outputs.  Additionally, each element can come equipped with a configuration input that can block or pass the inhibition inputs.  When the inhibition signals can pass, the two adjacent elements (and their corresponding variable-rate SPADs) competitively participate as a single Potts node;  when the inhibition signals are blocked, the two adjacent elements (and all elements to each side) do not compete; this can effectively convert a string of $q$ latching elements into two strings of smaller size.  In this way a bank of SPADs and latching elements may be configured into any number of CTMC neural configurations, allowing a single Potts annealer to operate with any combination of Potts model node sizes.

As an asynchronous circuit with potentially many stages of gates, two concerns arise regarding timing:  the behavior under coincident pulse inputs and effects on the annealing algorithm as the latch is transitioning and the one-hot constraint is momentarily not met.  In the first case, near-coincident pulse inputs have the capability to leave the one-hot latch with no high outputs.  This would occur if the falling edges of two pulses were coincident, leaving an inhibitory signal in all elements but no excitation signals.  Since pulse events are stochastic by design, it is inevitable that this will happen sometimes.  Similarly, delay in the inhibitory signal means that there will usually be brief instances during a normal transition during which multiple outputs are hot.  Both of these phenomena will be felt at the algorithmic level as intermittent inaccuracy of the energy calculations, which may or may not be problematic depending on the optimization problem.  In general however the incorrect energy calculation will be minimal: if a CTMC neuron changes state at roughly 100 MHz, and the delay through each latching stage is only 10 or 20 ps, a $q=10$ latch would be 100x faster than its input and would spend a similarly small portion of time in an incorrect state.

\pagebreak
\section{Mapping an Ising model to a Potts model}

During testing we matched a single theoretical distribution to the samples from both the Ising and Potts Boltzmann machines.  This required mapping from an eight-neuron Ising model to a Potts model with four four-state neurons. Ising states, biases, and weights will be denoted as $In$, $Ih$, and $Iw$ respectively; the Potts states, biases, and weights will be $Pn$, $Ph$, and $Pw$.  Each state of each Potts neuron is assigned a pair of states from two Ising neurons, shown in table \ref{table:1}.  The bias of each Potts state is the energy that state represents intrinsically, without interacting with other neurons in the Potts model; it includes the biases of the two Ising neurons that it is representing, and also the interaction energy between those two Ising neurons since they are subsumed within the single Potts neuron, also shown in table \ref{table:1}.

\begin{center}
\begin{tabular}{ c c c c }
 $Pn_i$ & $In_{2i-1}$ & $In_{2i}$ & $Ph_i(Pn_i)$ \\
 \hline
 1 & -1 & -1 & $-Ih_{2i-1}-Ih_{2i}+Iw_{2i-1,2i}$\\  
 2 & -1 & +1  & $-Ih_{2i-1}+Ih_{2i}-Iw_{2i-1,2i}$\\  
 3 & +1 & -1  & $+Ih_{2i-1}-Ih_{2i}-Iw_{2i-1,2i}$\\  
 4 & +1 & +1  & $+Ih_{2i-1}+Ih_{2i}+Iw_{2i-1,2i}$
\end{tabular}
\captionof{table}{Defining one Potts neuron based on two Ising neurons.}\label{table:1}%
\end{center}

Since a weight between two Potts neurons represents the interaction between four Ising neurons, there are four Ising weights that each Potts weight must account for (the two additional Ising weights found in a network of four Ising neurons are accounted for in the biases in table \ref{table:1}). Thus when mapping from the Ising model to the Potts model, each Potts weight is based on four terms, 

\begin{equation}
    Pw_{i,j}(P_i, P_j) = \pm Iw_{2i-1,2j-1} \pm Iw_{2i-1,2j} \pm Iw_{2i,2j-1} \pm Iw_{2i,2j}
\end{equation}

The sign of each term is determined by the corresponding signs of $In_{2i-1}$, $In_{2i}$, $In_{2j-1}$ and $In_{2j}$.  There are 16 combinations of signs, and 16 weight values connecting Potts neurons $i$ and $j$.

\pagebreak
\section{Hardware Implementation}

\begin{figure}[!t]
\begin{center}
    \begin{subfigure}[b]{6.5in}
        \centering
        \includegraphics[width=4.5in]{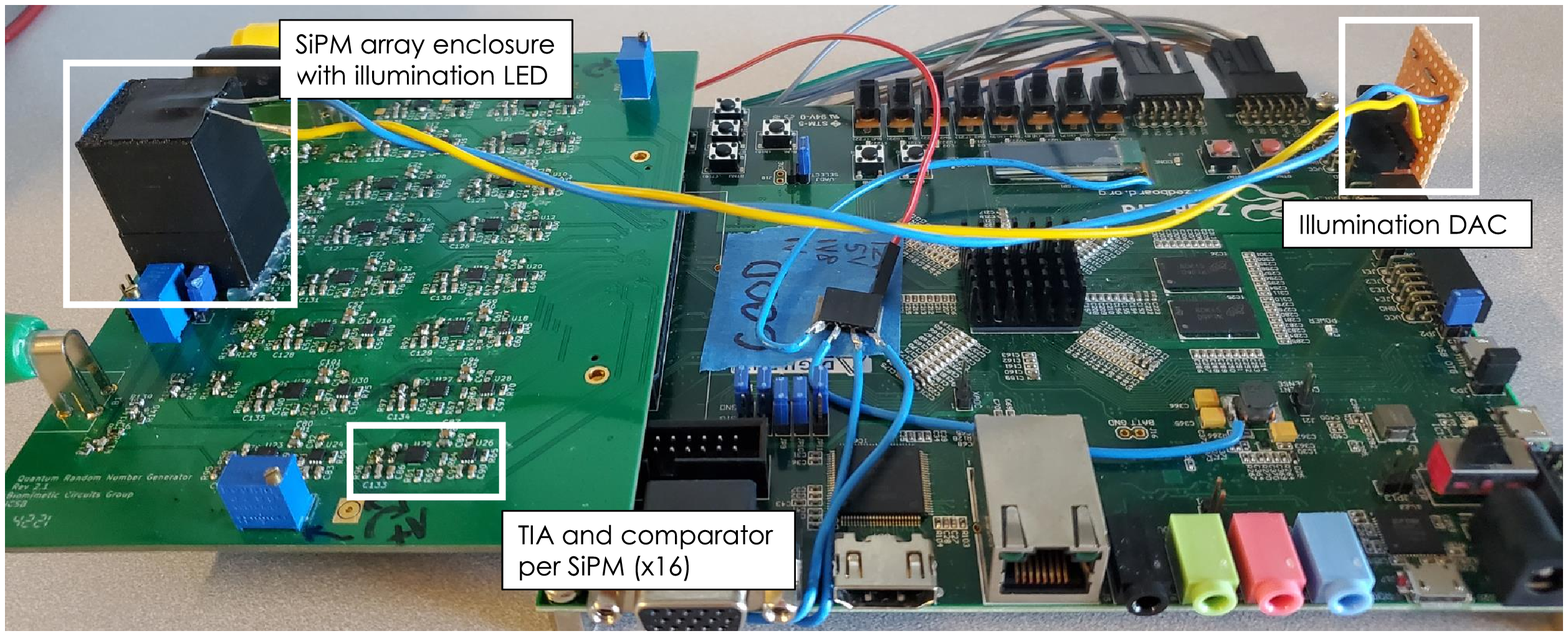}
    \end{subfigure}
    
    \vspace{0.1in}
    
    \begin{subfigure}[b]{6.5in}
        \centering
        \includegraphics[width=4.5in]{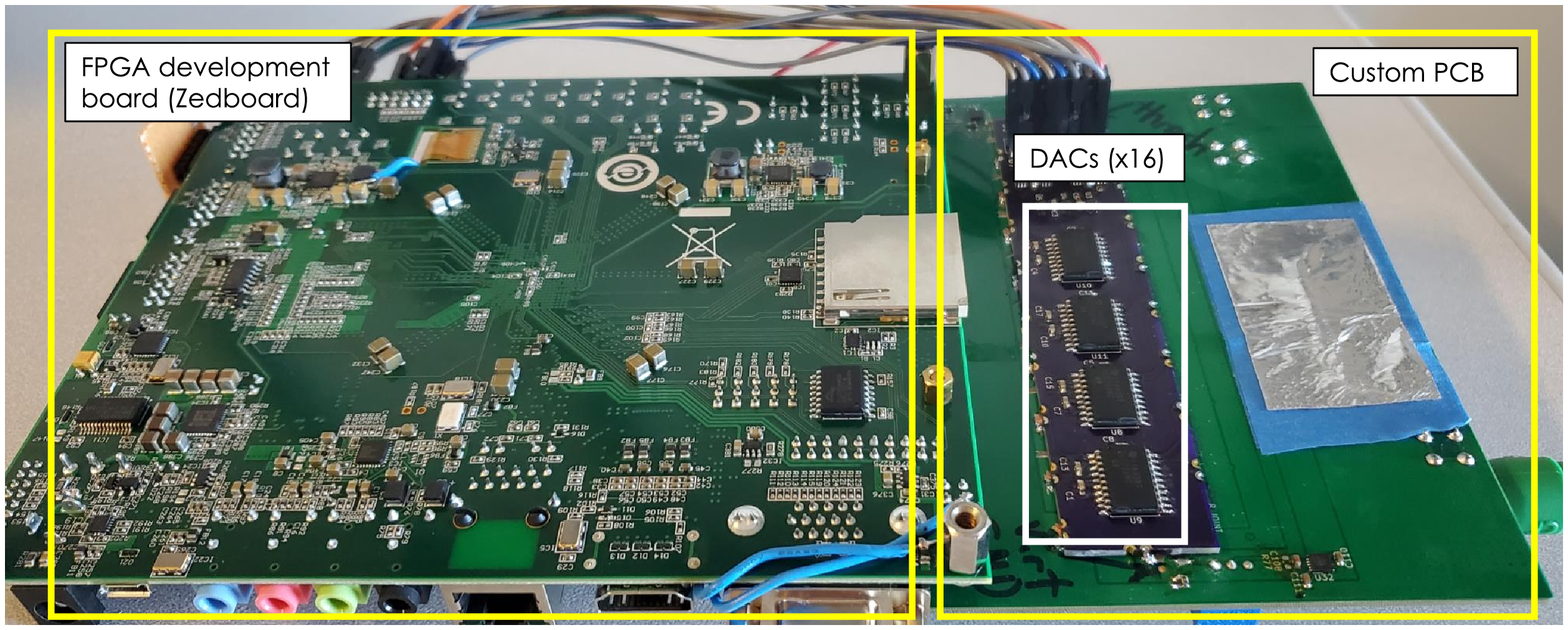}
    \end{subfigure}
\end{center}
\caption 
{ \label{fig:test_pcb}
Hardware for demonstrating SPAD CTMC neurons.} 
\end{figure}

All experimental data demonstrating SPAD CTMC neurons was collected using the hardware setup shown in figure \ref{fig:test_pcb}.  The setup consisted of 16 variable-rate SPAD circuits implemented on a custom PCB and an FPGA development board (Zedboard).  Asynchronous pulses from the variable rate SPADs (ON Semi ARRAYJ-30020-16P-PCB, using 16x LMH6629 for transimpedance sense amplifiers) were captured by the FPGA and used to control either binary RS latches (forming Ising neurons) or four-way latches (forming Potts neurons).  The latch states controlled adder trees inside the FPGA which calculated values (energies) to be written to the DAC of each variable-rate SPAD.  Due to the evolutionary nature of research, the DACs exist on an add-on PCB attached to the back of the main PCB.  Each time one of the neural state latches changed, all 16 DACs were updated; since the DACs could only be written over a serial interface (4x AD7226; 8-bit parallel, 50 MHz, 16 cycles to update all 16 DAC outputs), SPAD pulses were ignored until the DAC update completed in order to maintain correct sampling statistics.  In a scaled CMOS implementation, this delay would not be present.  The DACs set the thresholds on comparators (16x MAX40026) in each variable-rate circuit, affecting the pulse rate of each SPAD and in turn controlling the distribution of latch states held in the FPGA, forming a Boltzmann machine.  The behavior of this system was studied using the accompanying processor on the Zedboard's ZYNQ 7020 SoC, which ran Ubuntu with the PYNQ software layer, allowing fast collection of large quantities of data from a Jupyter Notebooks interface.  Illumination of the SiPM array was controlled by driving an LED with a DAC; the LED light was attenuated before reaching the SiPMs, and the scale of the illumination intensity was not measured. Thus the illumination intensity is only reported in arbitrary units (corresponding to the value set on the DAC).

\pagebreak
\section{Sampling methods}

Many measurements of the behavior of CTMC neurons, from the transfer function of single variable-rate SPADs to distributions of complete Ising and Potts annealers, required statistical sampling.  In general enough samples were accumulated such that deviations from ideal behavior could be attributed to the system itself rather than the distribution of the sample mean.

\subsection{Variable-rate SPAD transfer functions}
Two different sampling methods were used to measure the threshold-to-pulse-rate transfer functions of variable rate SPADs; both methods gave the same results.  The first method, which was only applied to the experimental variable-rate SPAD circuits, was to directly count the pulse edges received from the comparators by the FPGA.  In this case a threshold was physically set using the DACs and the resulting pulse edges were counted over a 100 ms time period by the FPGA; this was done iteratively over the full range of the 8-bit DACs, yielding 256-point transfer functions.

The second method was a computational analysis of the filtered SPAD outputs shown in figures (xx), obtained either from simulation or directly measured from the transimpedence amplifier using an oscilloscope.  In this method the comparator component of the variable-rate SPADs was performed computationally: using a single trace, positive crossings of a set threshold were counted, and the threshold was iteratively set to different values to measure a transfer function.  Performing this measurement on raw traces yielded higher event rates than (but the same exponential relationship as) the first method; this discrepancy was due to high-frequency sequential threshold crossings that were filtered out by the actual hardware.  Performing a low-pass filtering operation before computationally measuring the variable-rate SPAD transfer functions brought the magnitude of the rates in line with those obtained directly from hardware. 

The computational approach to measuring transfer functions was required especially for the simulation circuit results.  Otherwise a few hundred nearly identical simulations, each with a different threshold, would have been required; this would have been computationally prohibitive.  Instead computational time was spent on simulating a longer single trace, so that the tail of the transfer function could be measured with greater accuracy.  

Table \ref{table:2} lists lists measurement conditions and associated sampling uncertainty for the various variable-rate SPAD transfer function measurements.  The uncertainty depends on how many pulses were recorded during the sampling period, with higher pulse rates resulting in lower uncertainty.  The table lists the lowest measured pulse rates and associated number of detected pulses, from which a worst-case relative measurement error is determined.  The measurement quantity (number of pulses over the measurement timeframe) is modeled as a Poisson distribution.

\begin{center}
\begin{tabular}{ c c c c c c}
 circuit & threshold & figures & sampling time & min. count @ rate & relative error\\
 \hline
simulated & computational & 5g, 5h & 242 $\mu s$ & 242 @ 1 MHz & 0.064 \\
experimental & computational & 5e, 5f & 10 ms & 100 @ 10 kHz & 0.100 \\
experimental & en vivo & 6 & 100 ms & 1K @ 10 kHz & 0.031

\end{tabular}
\captionof{table}{Relative sampling error in variable-rate SPAD transfer function measurements}\label{table:2}%
\end{center}

\subsection{Tanh Ising neuron transfer curves}
The transfer function of Ising neurons, which were only measured for the experimental demonstration of SPAD CTMC neurons, was directly done en vivo by sweeping the biases $h_i$ of each neuron while leaving the weights $w_ij$ unused.  This was done after calibrating the rates of the individual variable-rate SPADs to be equal (by introducing bias offsets to each SPAD circuit).  With a set $h_i$ bias $10^5$ samples of the neuron states were collected, and the fraction of samples in the $+1$ state were tabulated for each Ising neuron.  This process was repeated for each integer bias value in the range $[-75, 75]$, representing the full usable input range of each neuron.  Since updates in a SPAD CTMC neuron are not clocked but occur whenever a SPAD circuit produces a pulse, samples were taken at a relatively slow 200 kHz so that successive samples were not too correlated.  Thus each point on the measured neuron transfer function represents the mean of $10^5$ roughly I.I.D. samples of a Bernoulli random variable, yielding a worst-case measurement uncertainty of $7x10^{-4}$ at neutral bias.

\subsection{Ising and Potts model distributions}
The distribution of samples produced by SPAD CTMC Ising and Potts models was measured by setting up a network configuration (weights, biases, and computational temperature) and then recording 5 million samples of the network state.  As with the measurement of individual Ising neuron transfer characteristics, consecutive samples are not independent.  For full Ising and Potts models the issue is even more acute, since it can take many updates for the collective model state to transition between regions of its state space; this is often referred to as the mixing problem, and its severity further depends on the particular configuration of weights and biases.  In addition, samples were recorded on a 50 MHz clock even though DAC update pausing (see Hardware Implementation) prevented the state from updating faster than 2 MHz.  In light of these obfuscations we present empirical evidence of measurement uncertainty instead of performing any rigorous bounding.  The 5 million samples were collected in 50 sub-batches, and after each batch the Kullback-Leibler  (KL) divergence between the measured distribution (cumulative across previously captured samples) and the theoretical distribution was calculated, shown in figure \ref{fig:kl_div}.  This figure shows that rather than continuing to decrease, the KL divergence plateaus, indicating that the difference between the measured and theoretical distributions is real and not an artifact of measurement uncertainty - otherwise the KL divergence would continue to decrease as more samples were recorded.  From this we can conclude that the measurement uncertainty of the Ising and Potts distributions is less than the visible deviations from the ideal distribution, and that, unsurprisingly, the SPAD-based Boltzmann machine does slightly deviate from ideal behavior.

\begin{center}
    \includegraphics[width=3.5in]{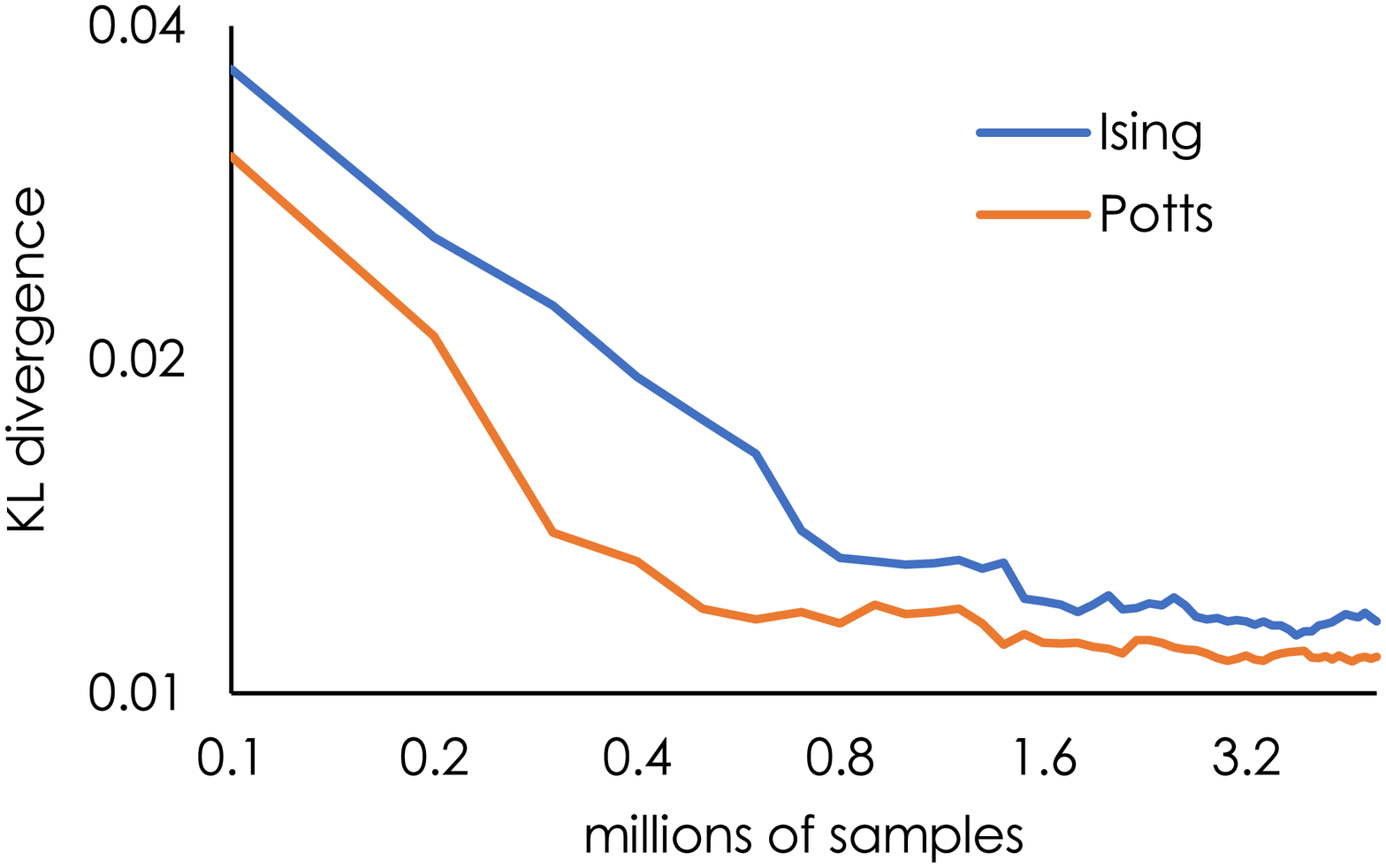}
    \captionof{figure}{Kullback-Leibler divergence of distributions sampled from Ising and Potts SPAD CTMC neurons, in reference to the theoretical distribution.}\label{fig:kl_div}%
\end{center}

%\pagebreak
%\section{Simulated Variable-rate SPAD}